\documentclass[preprint,
 amsmath,amssymb,
 aps]{revtex4-1}

\usepackage{graphicx}
\usepackage{color}
\usepackage{graphics}
\usepackage{float}
\usepackage{dcolumn}
\usepackage{bm}
\usepackage{url}
\usepackage{hyperref}

\graphicspath{{fig/}}

\usepackage{caption}
\usepackage{subcaption}

\begin{document}

%\preprint{}

\title{Parameter Estimation in Models with Complex Dynamics}

\author{Abhirup Ghosh}
 \email{abhirup.ghosh@icts.res.in}
\affiliation{International Centre for Theoretical Sciences (ICTS-TIFR), Bangalore 560089, India}%
\author{Samit Bhattacharyya}
 \email{samit.b@snu.edu.in}
\affiliation{Shiv Nadar University, Gautam Buddha Nagar, UP-208314, India }%
\author{Somdatta Sinha}%
 \email{ssinha@iisermohali.ac.in}
\affiliation{Indian Institute of Science Education and Research Mohali (IISER Mohali), Punjab-140306, India}%
\author{Amit Apte}%
 \email{apte@icts.res.in}
\affiliation{International Centre for Theoretical Sciences (ICTS-TIFR), Bangalore-560089, India}%

\date{\today}

\begin{abstract}
Mathematical models of real life phenomena are highly nonlinear involving multiple parameters and often exhibiting complex dynamics. Experimental data sets are typically small and noisy, rendering estimation of parameters from such data unreliable and difficult. This paper presents a study of the Bayesian posterior distribution for unknown parameters of two chaotic discrete dynamical systems conditioned on observations of the system. The study shows how the qualitative properties of the posterior are affected by the intrinsic noise present in the data, the representation of this noise in the parameter estimation process, and the length of the data-set. The results indicate that increasing length of dataset does not significantly increase the precision of the estimate, and this is true for both periodic and chaotic data. On the other hand, increasing precision of the measurements leads to significant increase in precision of the estimated parameter in case of periodic data, but not in the case of chaotic data. These results are highly useful in designing laboratory and field-based studies in biology in general, and ecology and conservation in particular.

\end{abstract}

\keywords{Parameter estimation; Bayesian methods; Ecological models; Population dynamics}
\maketitle

\textbf{Biological processes are often described by nonlinear
  mathematical models involving multiple parameters. Estimating these
  parameters is a common problem of data analysis, but its importance
  is compounded if the data-sets are short and noisy, and the models
  show bifurcations and chaos, as is quite common. In this paper, we
  show how, by adopting a simple Bayesian framework, one can infer
  crucial information about the underlying mathematical model, and
  provide bounds on the parameters describing them. The framework can
  be applied to any general chaotic dynamical system, as we illustrate
  through two examples: the logistic map as well as a mathematical
  model describing a real biological process, i.e., the population
  dynamics of \textit{Drosophila melanogaster}. We show how the
  posterior distribution for the parameters helps us discern the rich
  information about the parameters of the model as encoded in the
  noisy observations of the system dynamics, including the the
  statistical correlation between the various parameters.}

\section{Introduction} \label{sec:intro}

Mathematical models of complex phenomena are typically nonlinear and invariably involve multiple
parameters. The estimation of these parameters using observed
data is an important step in validating the models and later utilizing
them for predictive studies and for enhancing our
understanding of the real systems being modeled~\cite{aguirre2009modeling,
hong2008model, van2013detection}. The problem of parameter estimation extends well beyond biology, for example into astrophysics~\cite{schafer2015framework, abbott2016properties, ghosh2016testing}, where Markhov Chain Monte Carlo methods are commonly used to understand the behaviour of dynamical systems~\cite{RobertC99, JASA93_443p1032, ApteH07, ApteJ08, kantas2015particle}. However, this has especially far-reachng consequences in describing real biological processes, as experimental data-sets are typically small and noisy, and mathematical models show a range of dynamical behaviour - from stable to chaotic~\cite{glass1988clocks, suguna1999minimal, singh2004role}.

The wide prevalence of short and noisy experimentally obtained data-sets 
requires careful modeling of the observational errors and provides 
a significant motivation for a probabilistic approach to the parameter 
estimation problem, which has received significant attention recently~\cite{coelho2011bayesian,
alkema2008bayesian, bettencourt2008real, calderhead2009accelerating,
vyshemirsky2008biobayes, golightly2008bayesian, gao2012bayesian,
de2002fitting, rasmussen2011inference, harmon1997markov,
calder2003incorporating, andrieu2010particle}. Further, quoting from~\cite[p.15]{Hung-MIT-report},
it is known that ``most measurements of the state of a system contain comparatively little 
information about the parameters of the system except for those iterates 
where the hyperbolic behavior of a system becomes degenerate'' making it 
difficult to get precise estimates of the parameter by using
deterministic or variational methods such as least squares methods
(\cite{baker1996inverting, kostelich1992problems, jaeger1996unbiased, voss2004nonlinear, hong2008model} 
and references therein) or synchronization based approaches
(\cite{abarbanel2009dynamical, alonso2015parameter,
sorrentino2009using, parlitz1996synchronization,
parlitz1996estimating, maybhate1999use, maybhate2000dynamic, amritkar2009estimating} and references therein),
without detailed prior information about the parameters to be
estimated.

The study of parameter estimation presented in this paper is motivated
by the two important aspects, mentioned above, in the context
of ecology - viz. (i) ecological time series depicting population 
dynamics obtained from both laboratory and field experiments are typically 
small and noisy~\cite{sugihara1990distinguishing}, and (ii) the dynamics
of population growth of organisms with non-overlapping generations are
modeled mathematically using discrete maps, which show
period-doubling bifurcations with respect to multiple
parameters~\cite{may1976simple}. The former aspect involves modeling of the observational errors
while both these aspects together make any deterministic estimation
either imprecise or unreliable or both. 

In this paper, we use Bayesian inference and the Markov chain Monte Carlo method~\cite{meyer2000bayesian, meyer2001fast, mcsharry1999better} on two
discrete-time population growth models with increasing mathematical
and biological complexity in order to identify the main difficulties
in estimating parameters with limited prior knowledge of these
parameters. Another major motivation for Bayesian approach is that the
deterministic (e.g. synchronization based) or variational
(least-squares) methods do not allow us to account for highly
nonlinear correlations between different parameters to be
estimated. In contrast, such nonlinear correlations are naturally
represented in the Bayesian posterior distribution for the parameters,
as can be seen from figure~\ref{fig:nonlin-corr}. The knowledge of
such nonlinear relationships between parameters would be naturally
useful in further predictive or diagnostic studies using the models. 
All these features are present in ecological and epidemiological time 
series data, which are important for assessment and prediction of population density.

The novelty of this study is the focus on analysing, in
detail, the effects of the bifurcations that generically occur as a
function of the parameters to be estimated. In particular, we focus on
period doubling bifurcations which are very common in population
dynamical models (and indeed in many other nonlinear dynamical
models), thus making the following conclusion very generally
applicable in other systems as well: in the presence of these
bifurcations, the estimation of parameters in the chaotic regime can
only be done quite imprecisely, as compared to the estimation in the
periodic regime even under fairly high noise strength. The paper is 
organised as follows. We introduce the mathematical notation in section~\ref{sec:mathframe}, 
and present the main numerical results in the subsequent 
sections~\ref{sec:logistic}-\ref{sec:popdyn} on the single parameter 
Logistic map and  a multi-parameter life history-based realistic model for 
insect population dynamics, and a discussion of the results 
in the last section~\ref{sec:discuss}.

\section{Parameter estimation: mathematical formalism and numerical
  framework} \label{sec:mathframe}

We consider deterministic dynamical models $M(x_n,\theta, I)$ on a state space $\mathcal{X}$, defined by
\begin{equation}
x_{n+1} = M(x_n,\theta, I) \,, \quad x_n \in \mathcal{X} \,,
\label{eq:model} \end{equation}
where $\theta = (\theta^1, \dots, \theta^s) \in \mathbb{R}^s$ is the set of $s$~parameters of the model and $I$ is any other information that goes into the construction of the model (e.g., values of some known parameters which do not need to be estimated). Given a time series of length $t$ of noisy and partial observations, $y_n$, defined by:
\begin{equation}
y_n = h(x_n) + \eta _n \,, \quad n=1,2,..,t \,,
\label{eq:noisydata}\end{equation}
where $h(x_n)$ is a function of $x_n$, and $\eta_n$ is the intrinsic
experimental noise, the main question we will aim to study is the estimation of parameters $\theta$ based on these observations (or measurements) $\mathcal{Y} := \{y_1, \dots, y_t\}$. In this paper, we will consider the case of $y_n \in \mathbb{R}$ so that $\mathcal{Y} \in \mathbb{R}^t$.

We will consider the Bayesian probabilistic formulation of this
question which takes into account \emph{(i)} any prior information about the
parameters $\theta$ through the prior (unconditional) distribution
$r(\theta | I)$, and \emph{(ii)} the likelihood $l(\mathcal{Y} | \theta, I)$,
i.e., the conditional distribution of the data given the
parameters \cite{ghasemi2011bayesian, liepe2014framework,
  kruschke2014doing}. Given the prior and likelihood, Bayes' theorem gives the posterior
probabilty distribution of $\theta$, given data $\mathcal{Y}$.
\begin{equation}
  p(\theta | \mathcal{Y}, I) = \frac{r(\theta | I)
    l(\mathcal{Y}|\theta , I)}{p(\mathcal{Y}|I)}
\label{eq:bayestheorem}\end{equation}
Essentially, the above posterior quantifies all the information about the parameters $\theta$ contained in the observations $\mathcal{Y}$ and this will be the main object of study in the rest of the paper.

We will assume that the measurement noise is Gaussian and uncorrelated in time, i.e., $\eta _n$ are independent and identically distributed (iid) random variables derived from a normal distribution $\mathcal{N}(0, \sigma_r^2)$. We will call $\sigma_r$ as the ''strength'' (more precisely of course the standard deviation) of the ``recovery noise,'' to be contrasted later with the actual noise present in the observations. We will also assume that the prior probability distribution on the parameters $\theta$ is uniform within physically motivated ranges $\theta^i \in [\theta^i_{min}, \theta^i_{max}]$ for $i = 1, \dots, d$. With these two assumptions, the expression for the posterior probability distribution, $p(\theta | \mathcal{Y}, I)$, simplifies to:
\begin{eqnarray}
p(\theta | \mathcal{Y}, I) 
&\propto & \exp \left[-\sum_{n=1}^t\frac{\| y_n - h(x_n)\|^2}{2\sigma _r^2} \right] \,, \label{eq:posterior}\\
&\textrm{for }& \theta^i \in [\theta^i_{min}, \theta^i_{max}] \textrm{ for } i = 1, \dots, d \,.
\nonumber \end{eqnarray}
We use the standard Metropolis-Hastings algorithm
\cite{robert2013monte, kruschke2014doing} to sample the above
posterior.

We use a slight variation of a setup commonly known in the data
assimilation literature as ``identical twin experiments''
(\cite[Chap.~7]{reich2015probabilistic} or \cite[Chap.~6]{KalnayBook}). This is achieved by choosing an initial
conditions $x_{inj}$ and parameters $\theta _{inj}$ (called ``injected parameter'') in order to
generate a trajectory
\begin{equation}
x_{n+1}' = M(x_n',\theta _{inj}, I) \,,
\label{eq:trajectory}\end{equation} 
with $x_0'=x_{inj}$ and $n = 0, \dots, t-1$. Then the simulated data
set is obtained as:
\begin{equation}
y_n = h(x_n') + \eta _{inj, n}
\label{eq:injnoise} \end{equation}
where $\eta _{inj}$ is a specific realization of an iid random variable derived from $\mathcal{N}(0, \sigma _{inj}^2)$. We call $\sigma{inj}$ as the strength (more precisely, the standard deviation) of the ``injected noise''.

During the parameter estimation, we use the dynamical model given by equation~\eqref{eq:model} in the definition of the posterior distribution given in equation~\eqref{eq:posterior}, where $\sigma_r$ in equation~\eqref{eq:posterior} may not be equal to $\sigma_{inj}$ used in equation~\eqref{eq:injnoise}. The main justification behind making this distinction is that the actual experimental or observational noise $\sigma_{inj}$ \emph{may not be known} to us, and thus we may choose a different noise $\sigma_r$ in the process of parameter estimation.

Using the above mathematical formalism, we explore the qualitative features of the posterior probability distributions, as we vary the following aspects of the problem: (a) injected noise, $\sigma _{inj}$, (b) recovery noise, $\sigma _r$, (c) length of the data set, $t$ and (d) choice of the injected parameter $\theta _{inj}$ for which the trajectory $\{x_n'\}$ is either periodic or chaotic. As mentioned earlier, the two ecological models of increasing complexity, both in terms of number of parameters, nonlinearity and number of variables involved, that we focus on are - the logistic map in section~\ref{sec:logistic} and a realistic insect population dynamics (N-W) model in section~\ref{sec:popdyn}.

\section{Logistic map} \label{sec:logistic}
\begin{figure}[t!]
\centering
\includegraphics[width=\linewidth]{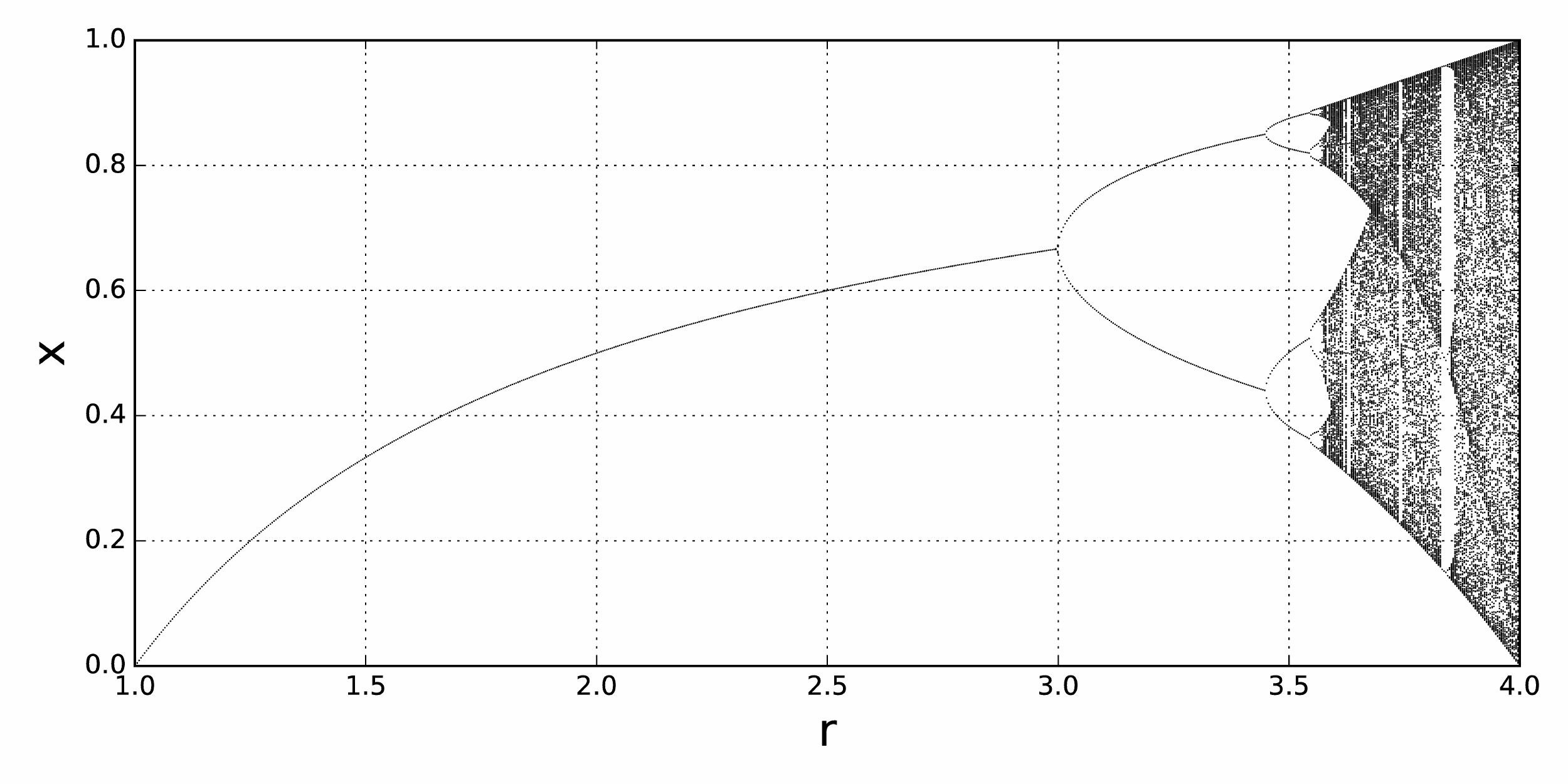}
\caption{\label{fig:bd_r}Bifurcation Diagram for logistic map}
\label{fig:logisticbif}
\end{figure}

\begin{figure}[t!]
\centering
\includegraphics[width=\linewidth]{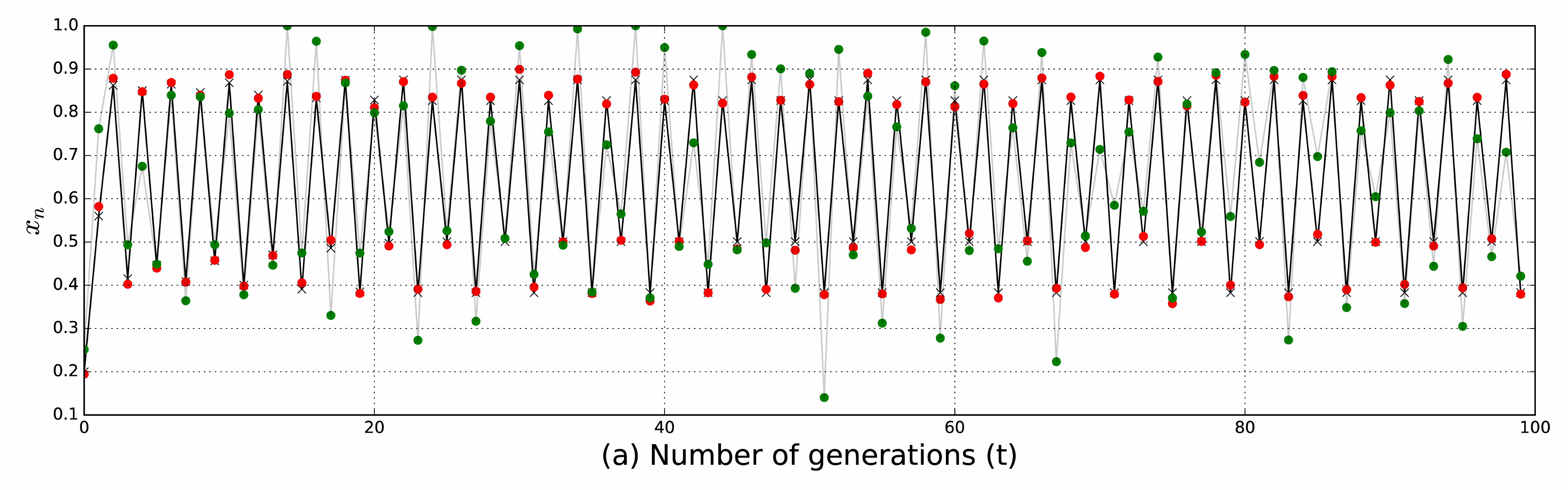}
\includegraphics[width=\linewidth]{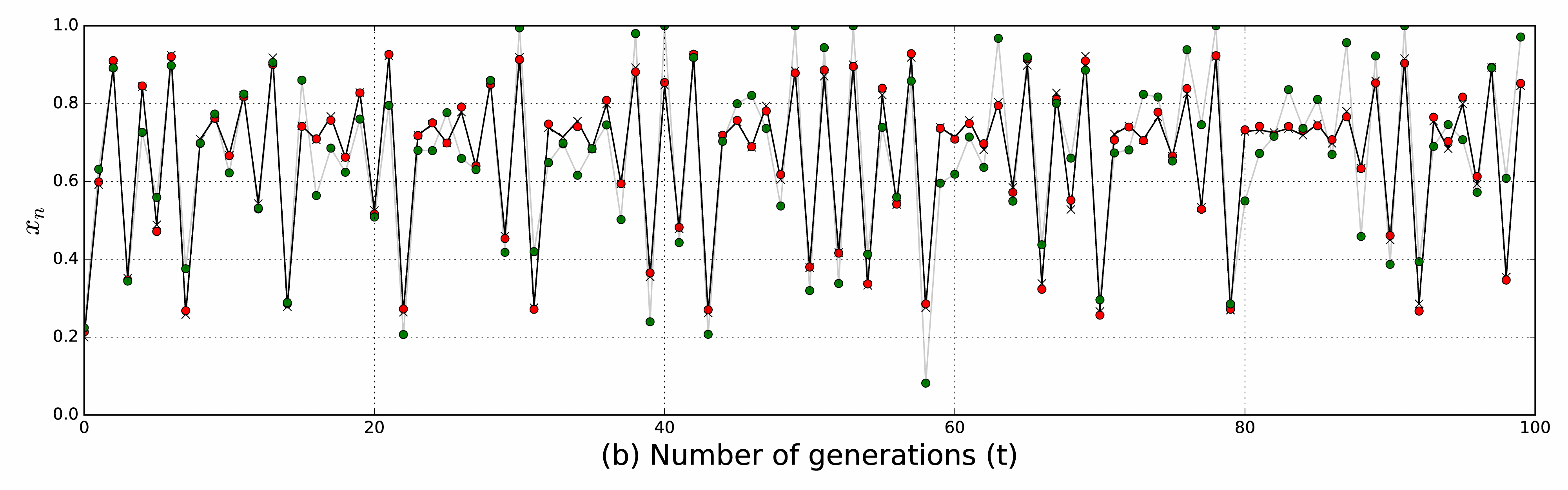}
\caption{\label{fig:data_lm}(a) Periodic data generated with $r=3.5$; (b) Chaotic data generated with $r=3.7$. In both cases the initial value $x_0$ is $0.2$. The solid black line and black crosses denotes the model trajectory $\{x_n\}$. The red dots represent the data $\mathcal{Y}$ for the smallest injected noise $\sigma_{inj} = 0.01$, and the green dots (connected by the grey line) represent data $\mathcal{Y}$ for the largest injected noise $\sigma_{inj} = 0.5$.}
\end{figure}

The logistic map is the simplest discrete one parameter population dynamic model, which is also the paradigm for studying period-doubling bifurcations and chaos in one-dimensional unimodal maps \cite{feigenbaum1978quantitative, may1976simple}.
\begin{eqnarray}
x_{n+1} = r x_n(1-x_n) \,, \qquad
x_n \in [0,1] \,, \ r \in [1,4] \,.
\end{eqnarray}
Here $x_n$ represents the population size (scaled by the carrying capacity, i.e., the maximum population size supported by the environment/habitat) at the $n^{th}$ generation. The bifurcation parameter, $r$, is the growth rate of the species. It is well known that this model exhibits a variety of dynamics with increasing growth rate, as shown by the well-known bifurcation diagram in Fig.~\ref{fig:logisticbif}. By appropriately choosing $r_{inj}$ from the bifurcation diagram in Fig.~\ref{fig:logisticbif}, the output $M(x_n',r_{inj})$, can be periodic or chaotic, giving rise to a periodic data series $\mathcal{Y}_p$ (Fig.~\ref{fig:data_lm}(a)) for $r_{inj} = 3.5$ or a chaotic data series $\mathcal{Y}_c$ (Fig.~\ref{fig:data_lm}(b)) for $r_{inj} = 3.7$ that we will use in the parameter estimation problem. For both figures, the solid black line and black crosses denotes the model trajectory $\{x_n\}$. The red dots represent the data $\mathcal{Y}$ for the smallest injected noise $\sigma_{inj} = 0.01$, and the green dots (connected by the grey line) represent data $\mathcal{Y}$ for the largest injected noise $\sigma_{inj} = 0.5$. We choose a broad prior - a uniform distribution over the interval $r \in [1,4]$.

%-----------------------------
%============================
% FIG 1: bifurcation diagram for r
%============================
%-----------------------------
%============================
% FIG 2: data vector for logistic map: periodic and chaotic
%============================
%-----------------------------

\begin{figure*}[t!]
\centering
\includegraphics[width=\linewidth]{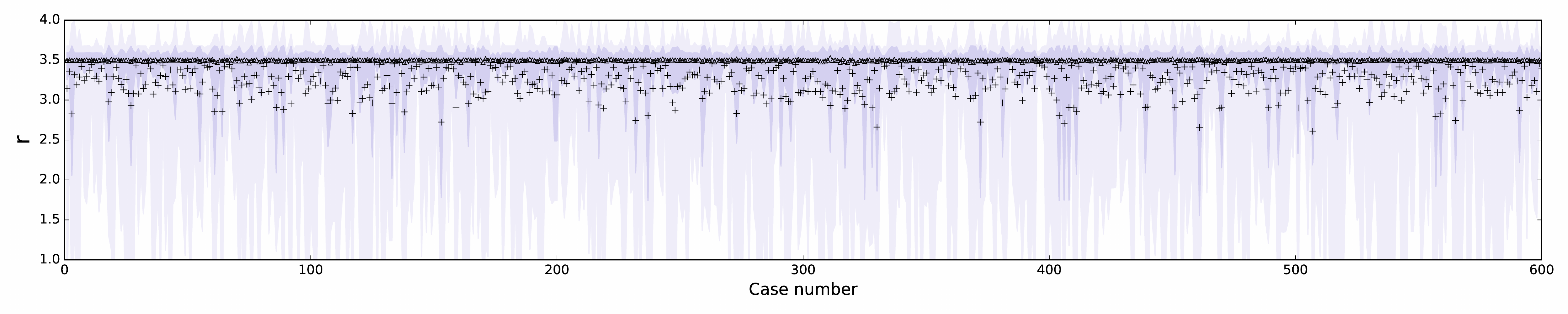}
\includegraphics[width=\linewidth]{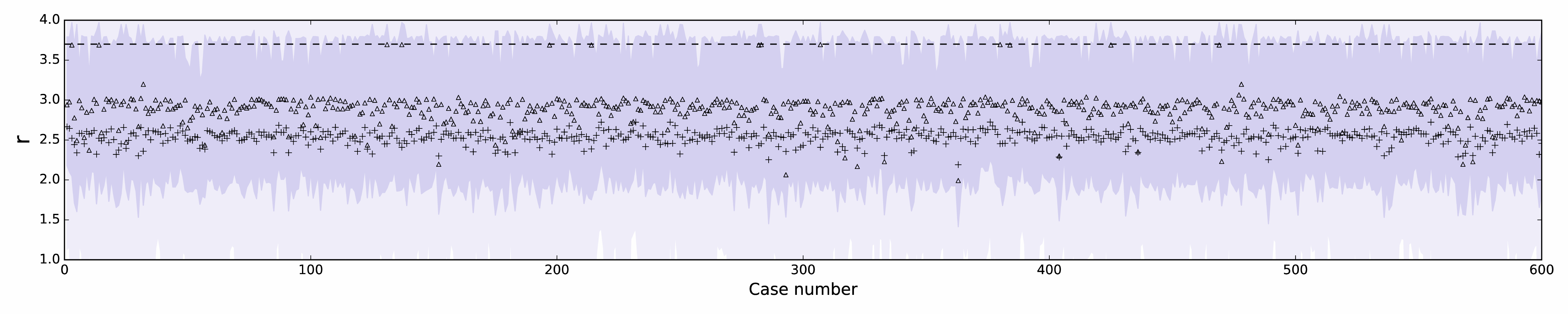}
\caption{\label{fig:robustness_lm}The mean (black plus), mode (black triangle), $68\%$ and $95\%$ confidence levels (dark and light slateblue bands) for the posterior for the parameter $r$ for each of the 600 different cases corresponding to injection value (black dashed line) of $r=3.5$ (for periodic data, top panel) and $r=3.7$ (for chaotic data, bottom panel), for a variety of values of length of data $t \in \{50, 100, 500, 1000, 5000 \}$, injected noise $\sigma_{inj} \in \{0.01, 0.02, 0.05, 0.1, 0.5\}$, recovered noise $\sigma_r \in \{ 0.01, 0.02, 0.05, 0.1, 0.5\}$, and initial condition of the Metropolis-Hastings $r_0 \in \{0.5, 1, 3, 3.5\}$ and $x_0 \in \{0.2, 0.8\}$.}
\end{figure*}

Using either the periodic or the chaotic data, we perform the standard Metropolis-Hastings sampling of the posterior $p(r | \mathcal{Y})$. In figure~\ref{fig:robustness_lm}, we show 1000 for periodic (top panel) and 600 different cases for chaotic data (bottom panel), respectively, corresponding to a range of values for $t$, $\sigma_{inj}$, $\sigma_r$. The red crosses in figure~\ref{fig:robustness_lm} show the mean, the blue crosses the mode, and the dark green or light green shading indicating the $68\%$ or $95\%$ confidence interval around the mean. We clearly notice that in the case of the periodic data, the mode is close to the injected value $r_{inj} = 3.5$ in many cases. However, neither the mean nor the mode is close to the injected value $r_{inj} = 3.7$ for the chaotic data. In fact, both the mean and the mode of $p(r | \mathcal{Y}_c)$ (for chaotic data) seem to be nearly independent of the values for $t$, $\sigma_{inj}$, $\sigma_r$, thus indicating that in absence of very detailed prior knowledge of the system (as quantified in the prior distribution), it is not possible to estimate parameters of a chaotic system. We now describe how the strength of injected noise $\sigma_{inj}$, the length of dataset $t$, and strength of observational noise $\sigma_r$ used during parameter estimation affect the posterior, both for periodic and chaotic data.

\begin{figure*}[t!]
\centering
\includegraphics[width=0.32\linewidth]{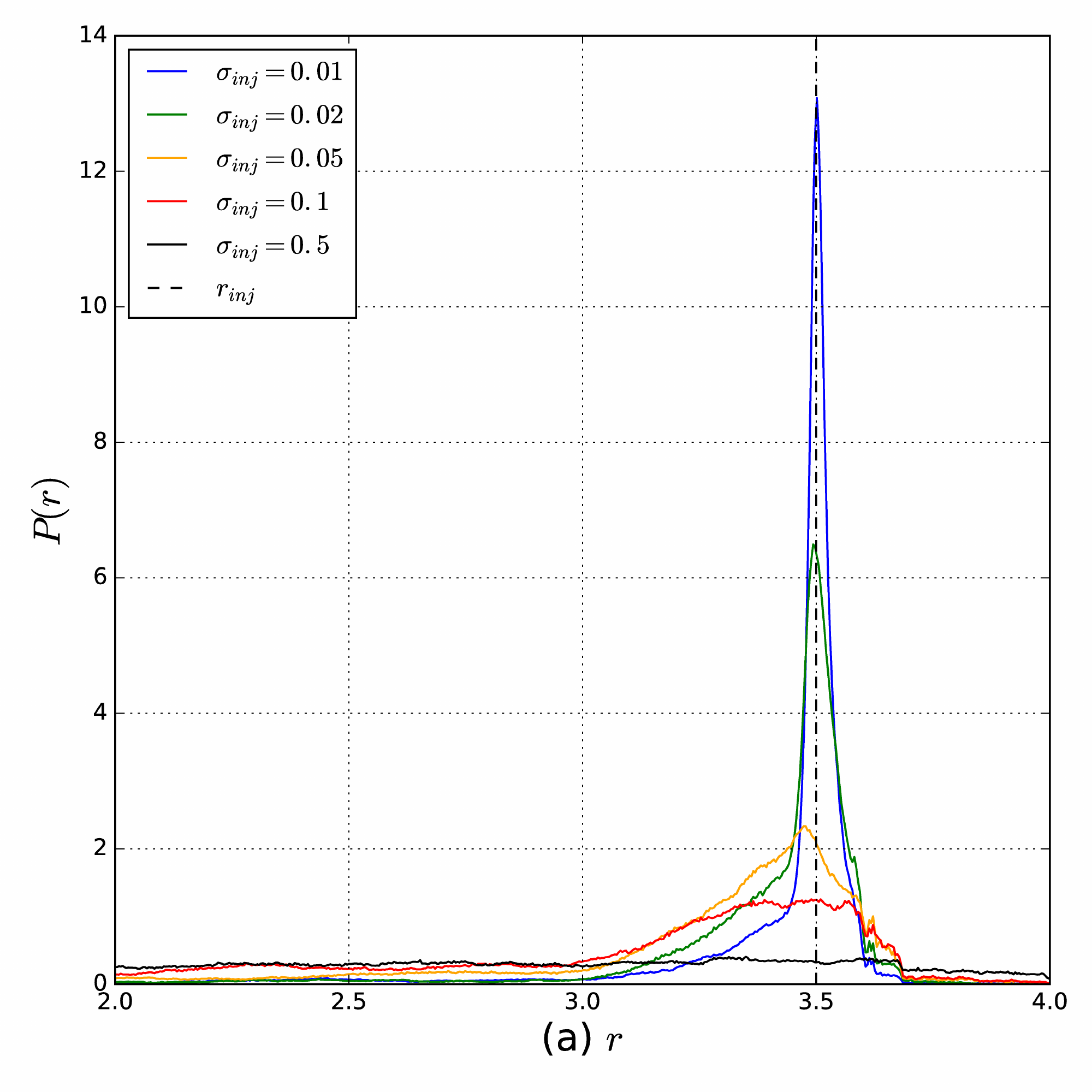}
\includegraphics[width=0.32\linewidth]{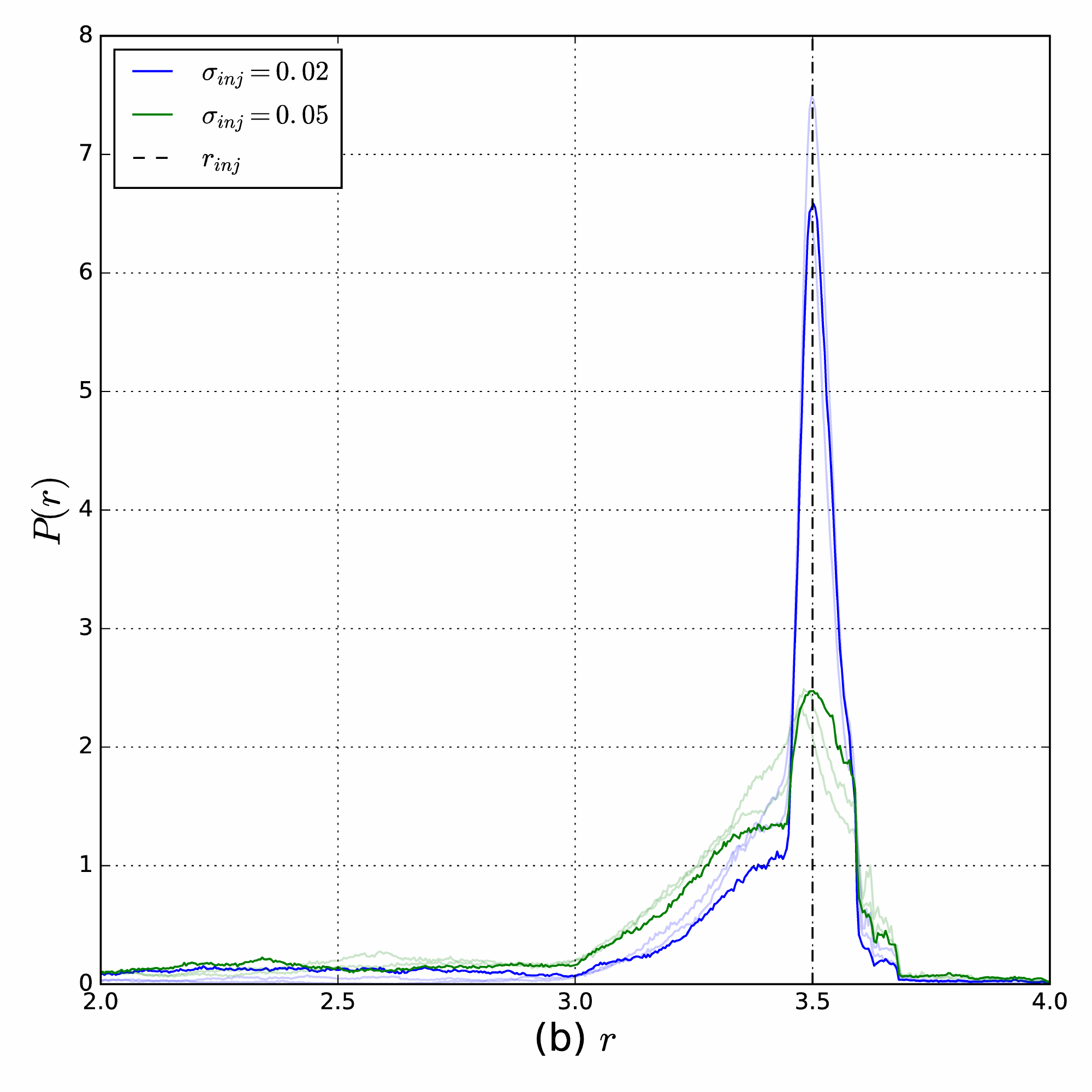}
\includegraphics[width=0.32\linewidth]{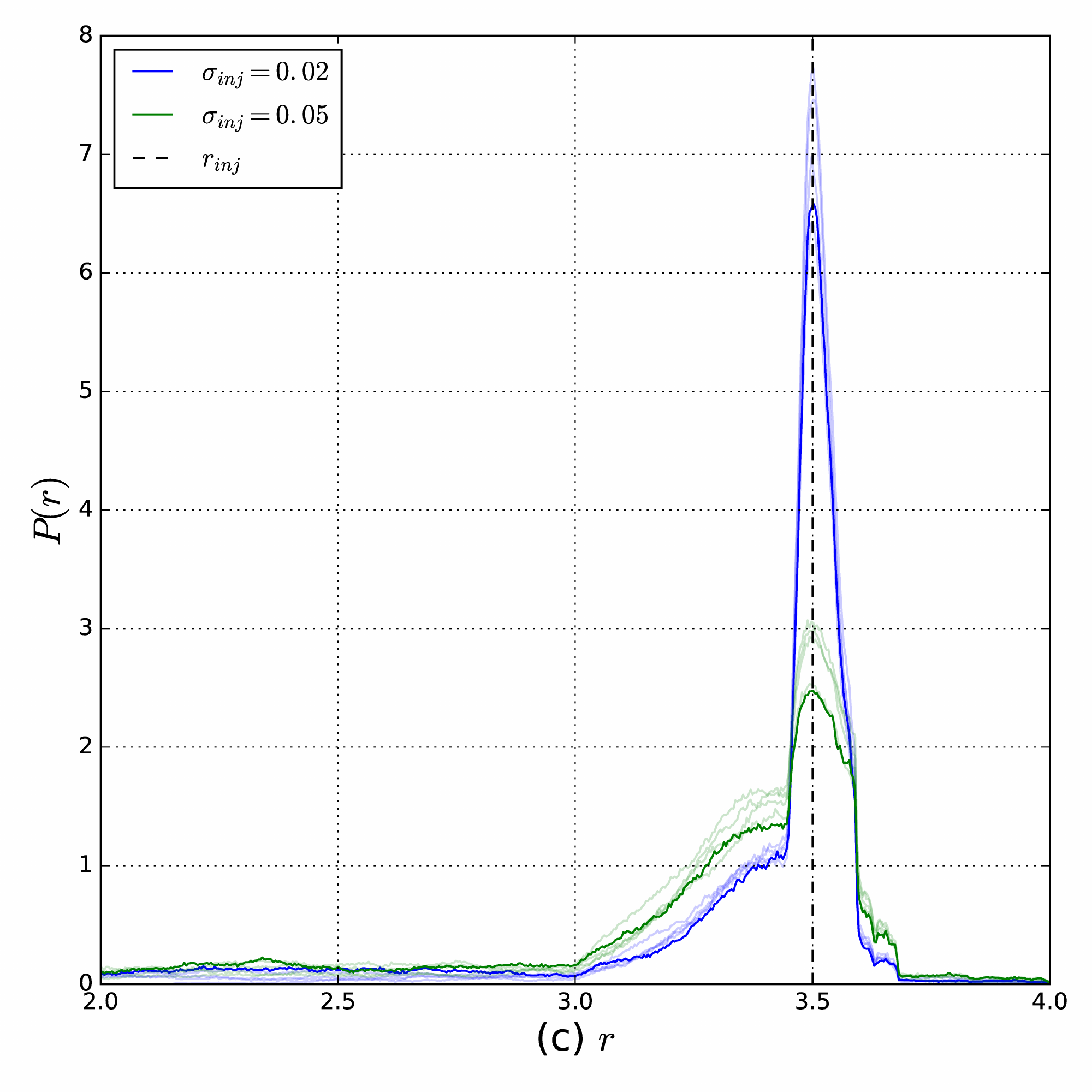}
\includegraphics[width=0.32\linewidth]{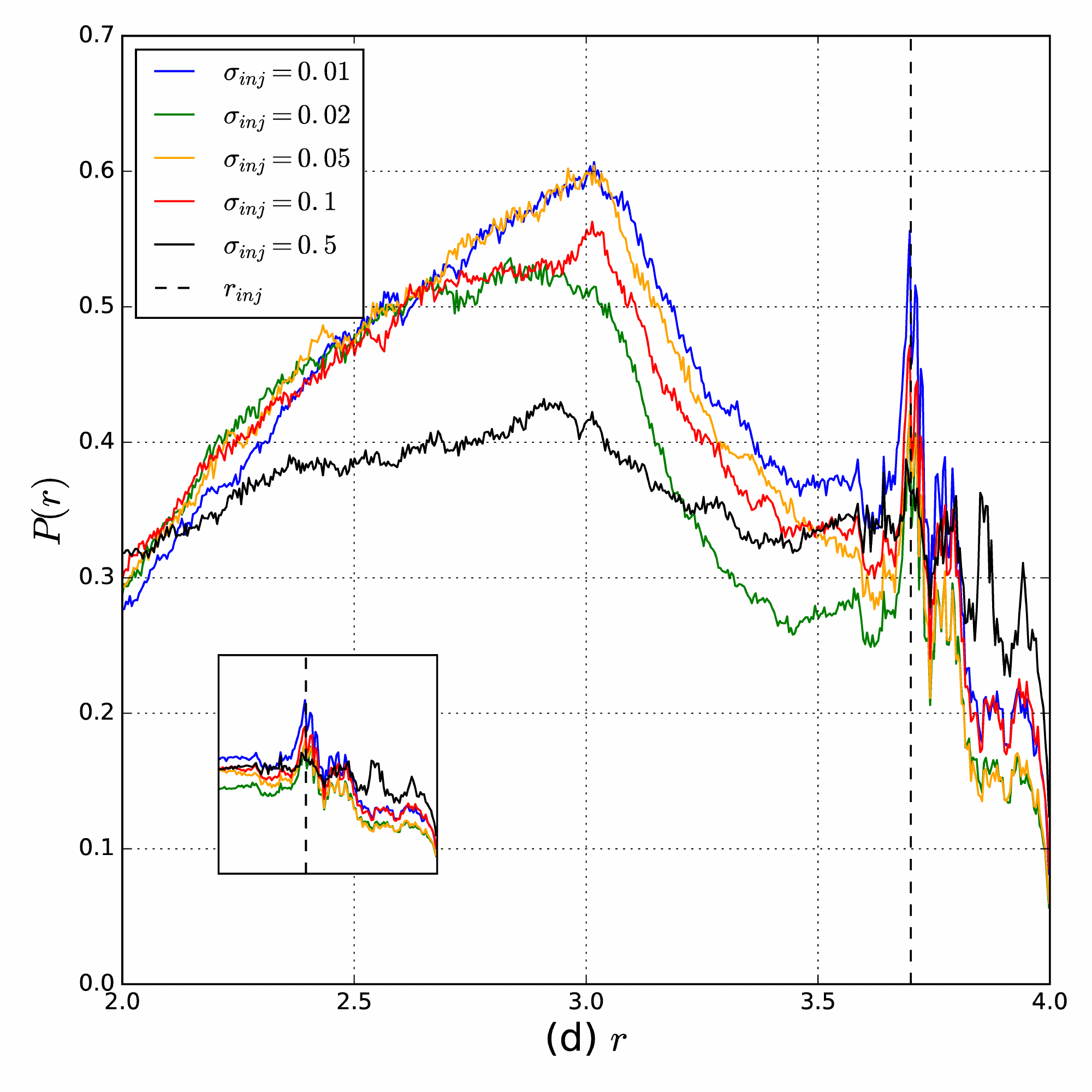}
\includegraphics[width=0.32\linewidth]{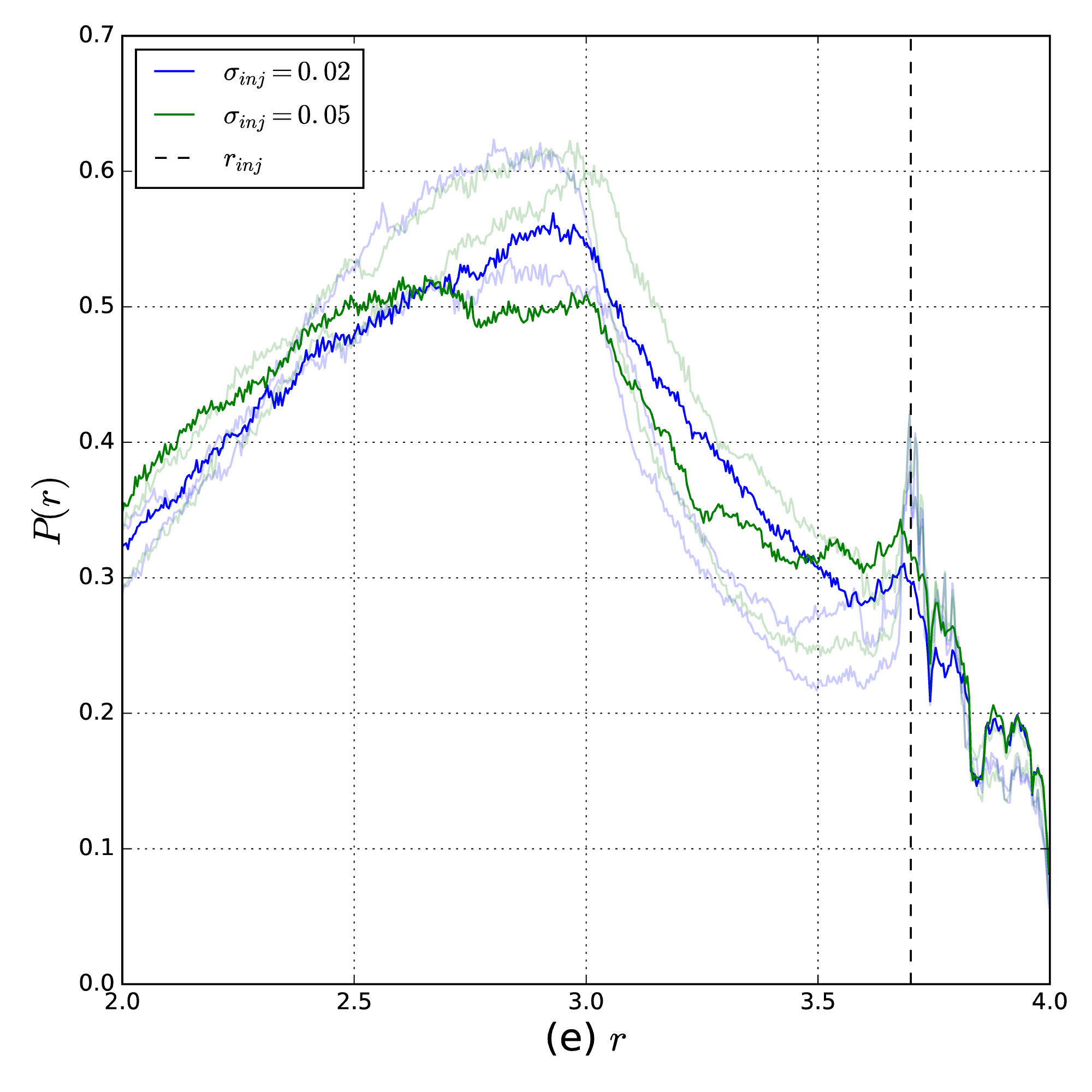}
\includegraphics[width=0.32\linewidth]{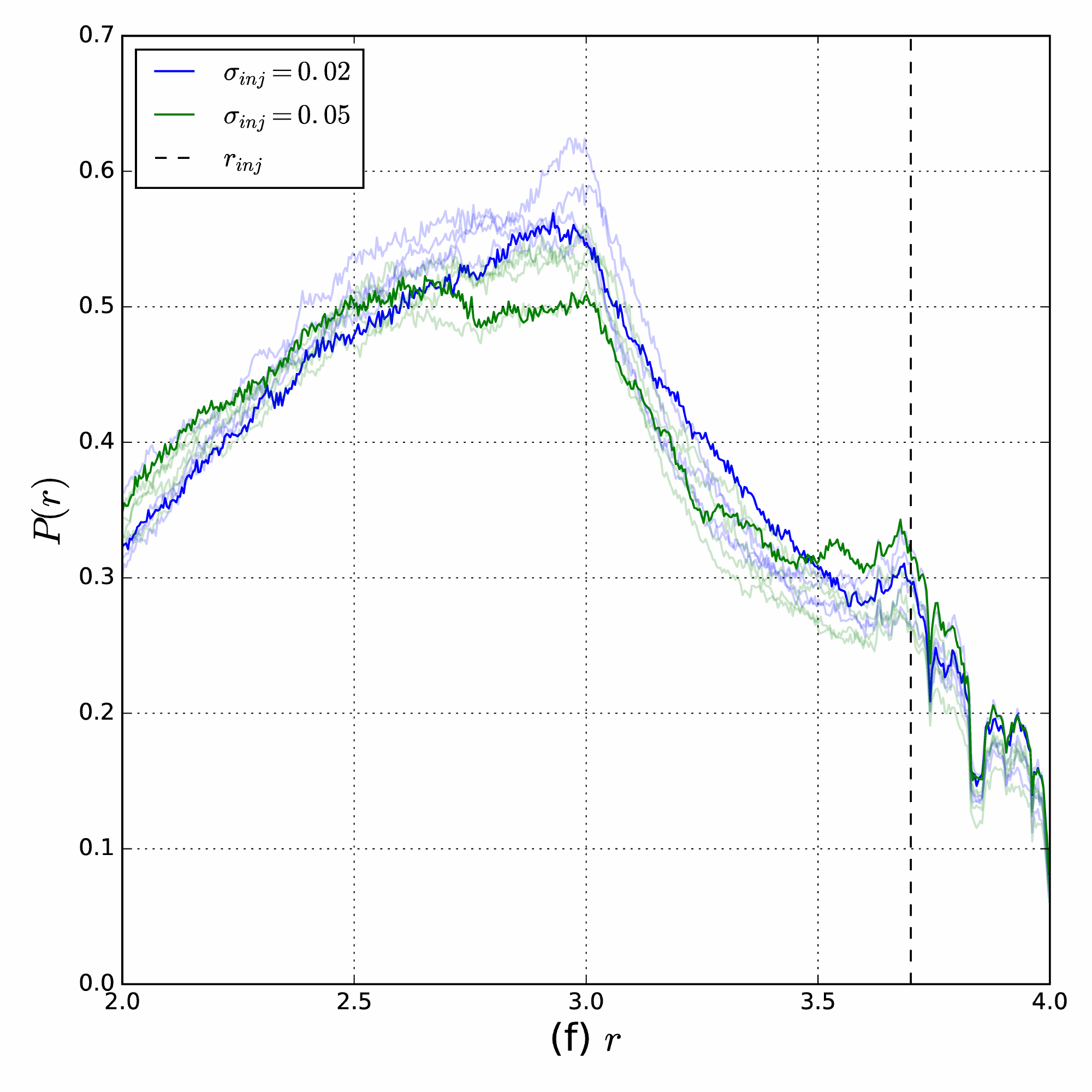}
\caption{\label{fig:lm_param_dependence}\emph{Top Row}: Posterior probability distributions for periodic data, $p(r | \mathcal{Y}_p)$. \emph{Bottom Row}: Posterior probability distributions for chaotic data, $p(r | \mathcal{Y}_c)$. \emph{Left Panel}: Posterior probability distributions for different levels of injected noise, $\sigma_{inj}$ = 0.01 (blue), 0.02 (green), 0.05 (orange), 0.1 (red), 0.5 (black). \emph{Middle Panel}: Posterior probability distributions for different lengths of data sets, $t=\{50,100,500\}$ for a given injected noise $\sigma_{inj}$ = 0.02 (blue) and 0.05 (green). The $t=500$ is solid, whereas the other curves are translucent. \emph{Right Panel}: Posterior probability distributions for different levels of recovered noise, $t=\{0.01, 0.02, 0.05, 0.1, 0.5\}$ for a given injected noise $\sigma_{inj}$ = 0.02 (blue) and 0.05 (green). The $t=500$ is solid, whereas the other curves are translucent. The inset in (d) shows a close up around the injected value of $3.7$ for chaotic data, giving rise to a secondary mode.}
\end{figure*}

All the plots in Fig.~\ref{fig:lm_param_dependence} show the posterior distribution for the logistic map, given in Eq.~\eqref{eq:posterior}, as sampled using the Metropolis-Hastings algorithm, for the periodic data $\mathcal{Y}_p$ in the top row (subplots (a), (b), (c)) and for the chaotic data $\mathcal{Y}_c$ in the bottom row (subplots (d), (e), (f)). The left column shows the effect of injected noise $\sigma_{inj}$, middle row that of length of dataset $t$, and the right column shows the posteriors for varying $\sigma_r$ for different values of $t$. Note the different ranges of the abscissa values $r$ in different plots.

\begin{figure}[t!]
\centering
\includegraphics[width=\linewidth]{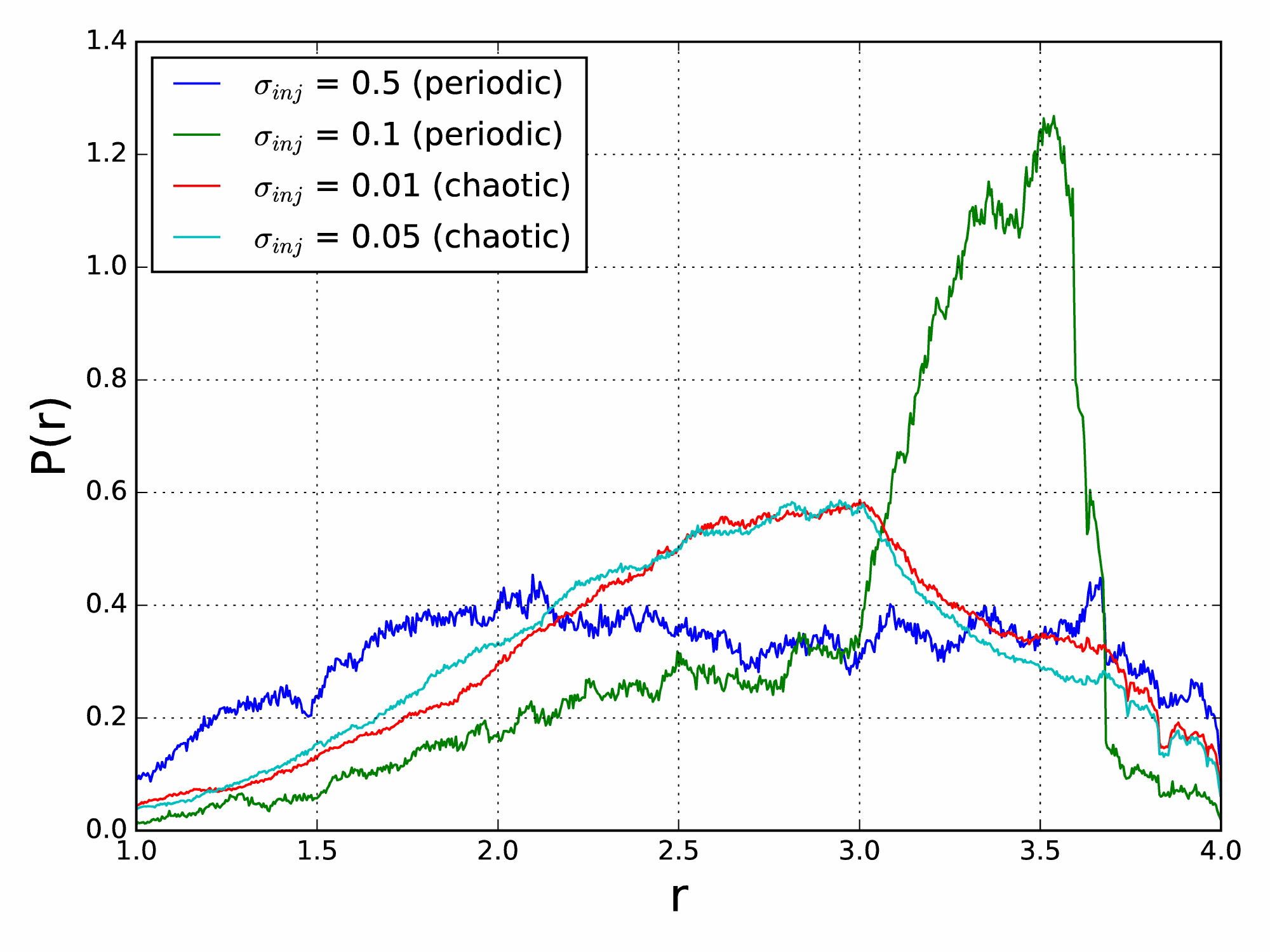}
\caption{\label{fig:lm_pe_vs_ch} Posterior probability distributions $p(r | \mathcal{Y})$ for periodic data with injected noise, $\sigma_{inj} = 0.5$ (blue), $\sigma_{inj} = 0.1$ (green) and for chaotic data with injected noise, $\sigma_{inj} = 0.01$ (red), $\sigma_{inj} = 0.05$ (cyan). Periodic data with large injected noise becomes indistinguishable from chaotic data with even very small injected noise.}
\end{figure}

Comparison of the three panels in Fig.~\ref{fig:lm_param_dependence}, the top row shows that for the case of parameter estimation with periodic data $\mathcal{Y}_p$, the injected noise $\sigma _{inj}$ has the most notable effect on the posterior $p(r | \mathcal{Y}_p)$ - more precise measurements lead to more precise parameter estimation [Fig.~\ref{fig:lm_param_dependence} (a)]. We also see that the effects of $t$ and $\sigma _r$ are negligible - varying either of these two does not lead to any major change in the posterior [Fig.~\ref{fig:lm_param_dependence} (b) and (c), respectively]. As expected, increasing $\sigma _{inj}$ leads to larger standard deviation of $p(r | \mathcal{Y}_p)$, to a point where parameter estimation on periodic data with high $\sigma _{inj}$ is indistinguishable from parameter estimation on chaotic data (as seen very clearly in Fig.~\ref{fig:lm_pe_vs_ch}).This is because, with increasing $\sigma _{inj}$, the underlying periodicity of the data is lost in the noise. 

The bottom row of Fig.~\ref{fig:lm_param_dependence} shows effect of these same parameters $\sigma_{inj}, t, \sigma_r$ on the posterior distribution conditioned on chaotic data. We see that the posterior is now much more broad as compared to the case of periodic data. In fact, the broad posterior distribution for parameter conditioned on chaotic data does not vary much with varying either $t$ or $\sigma_{inj}$ or $\sigma_r$. Fig.~\ref{fig:lm_param_dependence}(e) highlights an interesting observation in the parameter estimation of chaotic data for different levels of injected noise. For low injected noise, we see a bimodal distribution, where one of the modes is around the injected value, $r_{inj}=3.7$. This secondary mode however becomes less and less prominent with increasing injected noise, as compared to the primary mode [Fig.~\ref{fig:lm_param_dependence}(e) inset]. This seems to indicate that in order to use chaotic data to estimate parameters, one certainly needs to use other information about the system, which needs to be codified in the form of a narrow prior distribution rather than a very broad prior as we have chosen.

%============================
% FIG 3: Robustness of MH algorithm
%============================
%-----------------------------

%============================
% FIG 4: logistic map parameter dependence
%============================
%-----------------------------

%============================
% FIG 5: periodic vs chaotic posteriors
%============================
%-----------------------------

%-----------------------------

\section{The multi-parameter (N-W) model of insects population growth} \label{sec:popdyn}

We now apply the parameter estimation method to a more realistic model involving multiple parameters and variables, which is more likely to be used to study population growth processes. Many organisms, like insects, go through multiple life stages (egg, larva, pupa, adult) before reaching adulthood to lay eggs again for starting the next generation. Based on the details of the life stages of insects, a coupled difference equation model (termed as \textbf{N-W} model), is developed for insects with adult number (N) and body weight (W) as the two dynamical variables  \cite{10.2307/2461536, doi:10.1086/284890, Mueller01061988, 10.2307/4835, sinhajoshi}:
	\begin{eqnarray}
		N_{t+1}&=&\bigg(\frac{Cf_{max}N_t}{2(1+dN_t)}\bigg) exp\bigg(-\frac{b* log(A/W_t)}{a}-\bigg(\frac{cf_{max}N_t}{2(1+dN_t)} exp\bigg[-\frac{b *log(A/W_t)}{a}\bigg]\bigg)\bigg)\\
		W_{t+1}&=&A exp\bigg(-\bigg(\frac{af_{max}N_t}{2(1+dN_t)} exp\bigg[-\frac{b* log(A/W_t)}{a}\bigg]\bigg)\bigg)
	\end{eqnarray}
% for revtex 2 column mode
% \begin{widetext}
% 	\begin{eqnarray}
% 		N_{t+1}&=&\bigg(\frac{Cf_{max}N_t}{2(1+dN_t)}\bigg) exp\bigg(-\frac{b* log(A/W_t)}{a}-\bigg(\frac{cf_{max}N_t}{2(1+dN_t)} exp\bigg[-\frac{b *log(A/W_t)}{a}\bigg]\bigg)\bigg)\\
% 		W_{t+1}&=&A exp\bigg(-\bigg(\frac{af_{max}N_t}{2(1+dN_t)} exp\bigg[-\frac{b* log(A/W_t)}{a}\bigg]\bigg)\bigg)
% 	\end{eqnarray}
% \end{widetext}
where ($N_t, W_t$) and ($N_{t+1}$, $W_{t+1}$) denote the number and weight of the adult population in consecutive generations. The equations define a two dimensional map of adult population number and their mean weight from one time point (generation) to the next. The parameters $\{C,c,f_{max},b,A,a,d\}$ form the life-history parameter vector $\vec{\theta}$. These parameters relate to the different nonlinear stage-specific density-dependent feedback processes regulating the adult number, fecundity, and weight.

In the notation set up in section~\ref{sec:mathframe}, the N-W model is the dynamical model $M(x_n,\theta, I)$, where $x_n=\{N,W\}$ and $\theta=\{C,c,f_{max},b,d,A,a\}$. While the parameters C, A, $f_{max}$, can be related to physically intuitive concepts, the sensitivity parameters a, b, c and d are rather abstract, and it is not possible to ascertain a lot about these parameters from the above equations, except that they should all be positive. Hence the main concern is to estimate the values of the parameters, or to even just get bounds on their values. The problem is compounded by the presence of intrinsic noise in the experiment, which makes standard parameter estimation methods inadequate to solve this problem, and favours a Bayesian formalism, where one aims to obtain a posterior distribution on the parameters for a given data set.

As a first step we start with the estimation of a single parameter, $d$. The remaining parameters $\{C,c,f_{max},b,A,a\}$ are fixed at $\{$0.8, 0.002, 80.429, 0.001, exp(0.592), 0.003$\}$. The parameter values are adopted from literature \cite{10.2307/2461536, doi:10.1086/284890, Mueller01061988, 10.2307/4835}. Thus our model can be simplified to $M(x_n,d, I)$, where $I=\{C,c,f_{max},b,A,a\}$. Later we also discuss estimation of multiple parameters, for example, $\theta = (d,b)$, in which case $I=\{C,c,f_{max},A,a\}$ fixed at the same values as listed above.

\begin{figure}[t!]
\centering
\includegraphics[width=\linewidth]{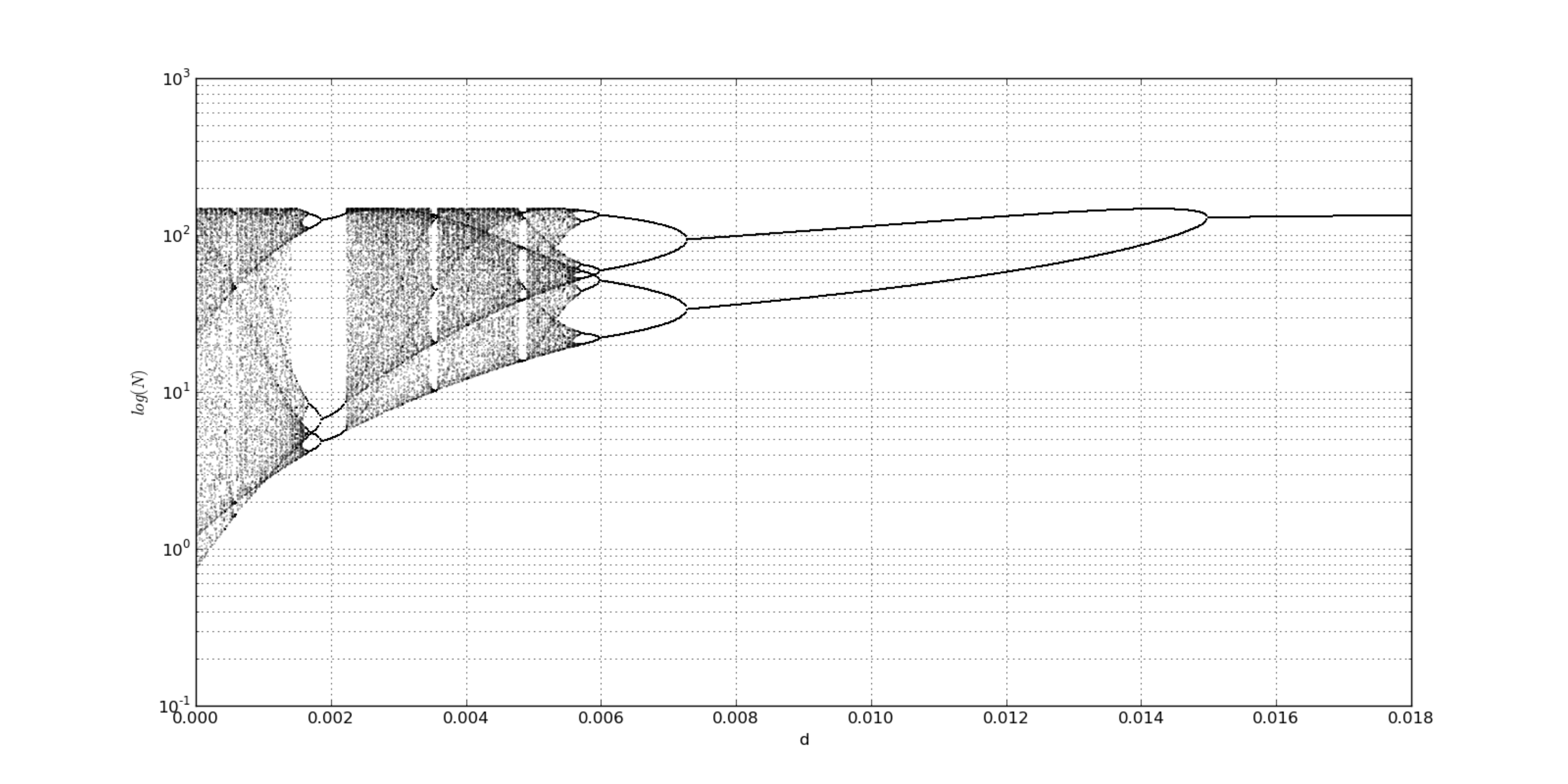}
\includegraphics[width=0.9\linewidth]{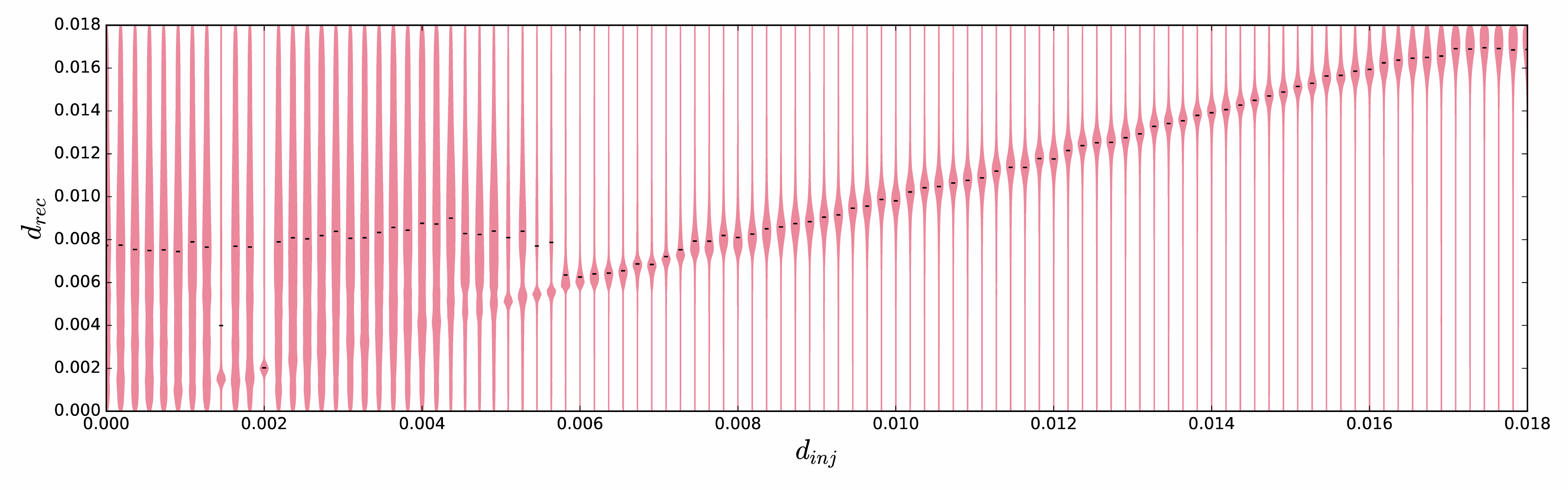}
\caption{\label{fig:bd_d}Bifurcation Diagram for $d$ (top panel). See text for other parameter values. The bottom panel shows a ``violin plot'' \cite{violin} with the injected value on the x-axis and the mean of the posterior on the y-axis.}
\end{figure}

\begin{figure}[t!]
\centering
\includegraphics[width=\linewidth]{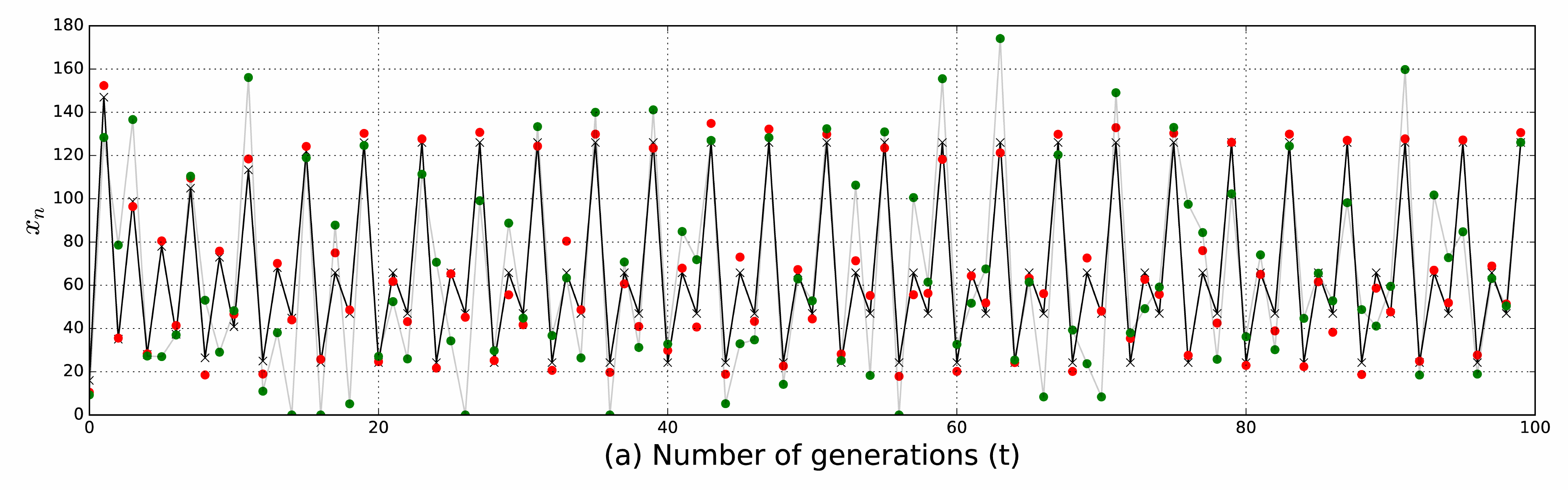}
\includegraphics[width=\linewidth]{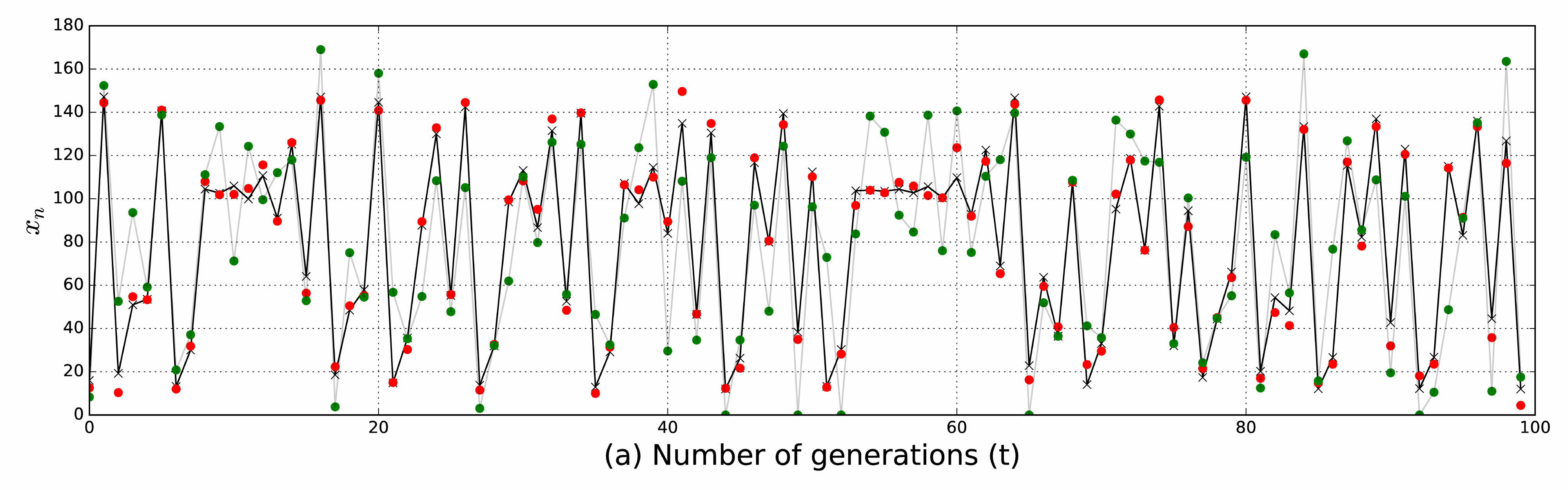}
\caption{\label{fig:data_droso}(a) Periodic data generated with $d=0.0065$; (b) Chaotic data generated with $d=0.004$. In both cases the initial value $n_0$ is $16.0$. The solid black line and black crosses denotes the model trajectory $\{x_n\}$. The red dots represent the data $\mathcal{Y}$ for the smallest injected noise $\sigma_{inj} = 5$, and the green dots (connected by the grey line) represent data $\mathcal{Y}$ for a large injected noise $\sigma_{inj} = 25$.}
\end{figure}

In both these cases of single or two parameter estimation, given a time series of observed populations (containing noise), $y_n \equiv N_{obs,n} \in \mathcal{Y}$, the likelihood function can be defined along the same lines as Eq.~\eqref{eq:trajectory}:
\begin{equation}
p(\mathcal{Y}|\theta , I) \propto  \exp \left[-\sum_{n=1}^t\frac{\| N_{obs,n} - N_{th,n}(\theta,I)\|^2}{2\sigma _r^2} \right]
\label{eq:droso-likelihood} \end{equation}
where, $N_{th,n}$ is the theoretical predictions of the model. Regarding data generation, in complete analogy with the logistic map case, an appropriate initial choice of $\theta_{inj}$ from the bifurcation diagram in Fig.~\ref{fig:bd_d}, outputs a periodic data series $\mathcal{Y}_p$ (Fig.~\ref{fig:data_droso}(a)), or a chaotic data series $\mathcal{Y}_c$ (Fig.~\ref{fig:data_droso}(b)). We then use this data series in the likelihood function defined in the above equation~\eqref{eq:droso-likelihood} and use the standard Metropolis-Hastings algorithm for sampling the posterior as defined in Eq.~\eqref{eq:posterior}. Since there is very little known \emph{a priori} about the parameters to be estimated, we again choose the prior which is uniform over physically motivated ranges.

%============================
% FIG 6: bifurcation diagram for d
%============================
%-----------------------------
%============================
% FIG 7: data vector for logistic map: periodic and chaotic
%============================
%-----------------------------

\subsection{Estimation of one parameter}
\begin{figure*}[t!]
\centering
\includegraphics[width=\linewidth]{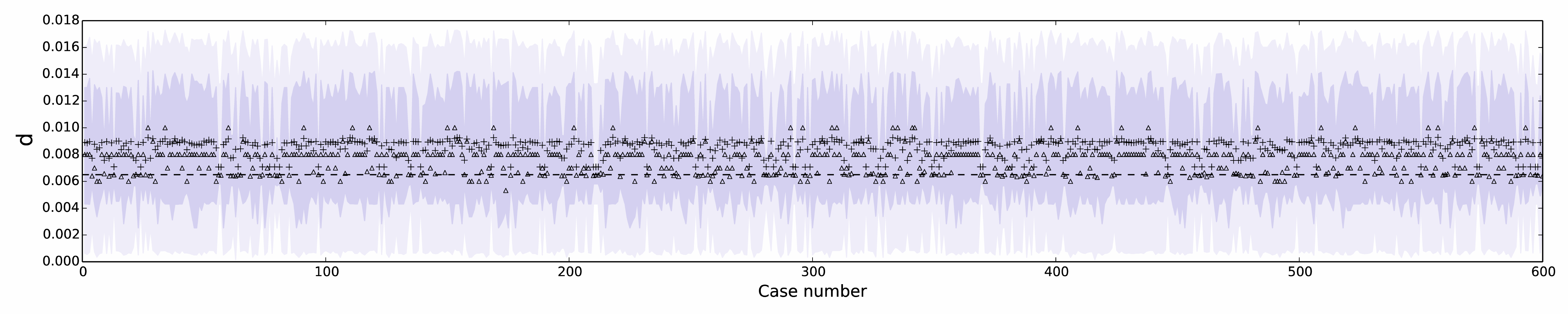}
\includegraphics[width=\linewidth]{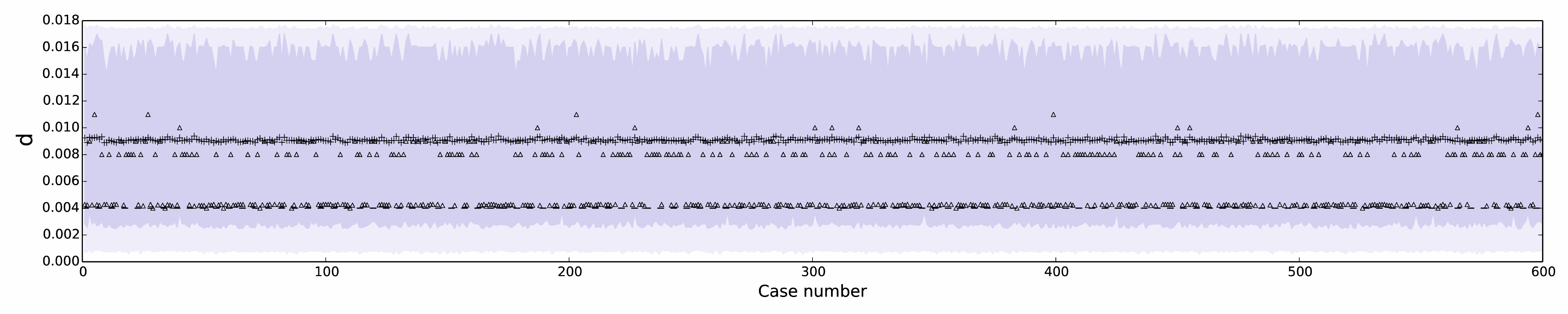}

\caption{\label{fig:robustness_droso}The mean (black plus), mode (black triangle), $68\%$ and $95\%$ confidence levels (dark and light slateblue band) for the posterior for the parameter $r$ for each of the 600 different case corresponding to injection value of $d=0.0065$ (for periodic data, top panel) and $d=0.004$ (for chaotic data, bottom panel), for a variety of values of length of data $t \in \{50, 100, 500, 1000\}$, injected noise $\sigma_{inj} \in \{5,10,25,50,75\}$, recovered noise $\sigma_r \in \{5,10,25,50,75\}$, and initial condition of the Metropolis-Hastings $d_0 \in \{0.001, 0.0065, 0.009, 0.016\}$ and $N_0 (10, 16)$.}
\end{figure*}

\begin{figure*}[t!]
\centering
\includegraphics[width=0.32\linewidth]{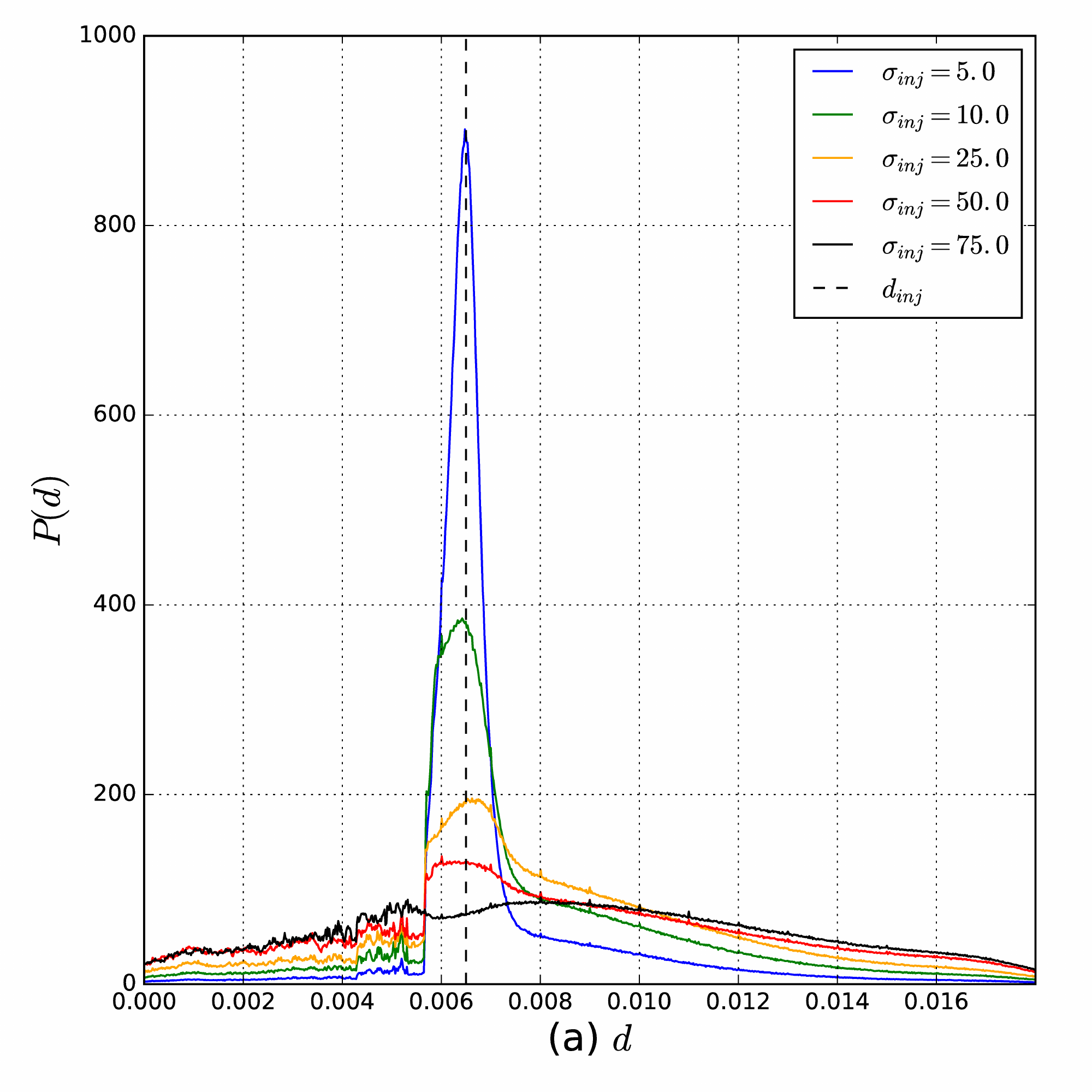}
\includegraphics[width=0.32\linewidth]{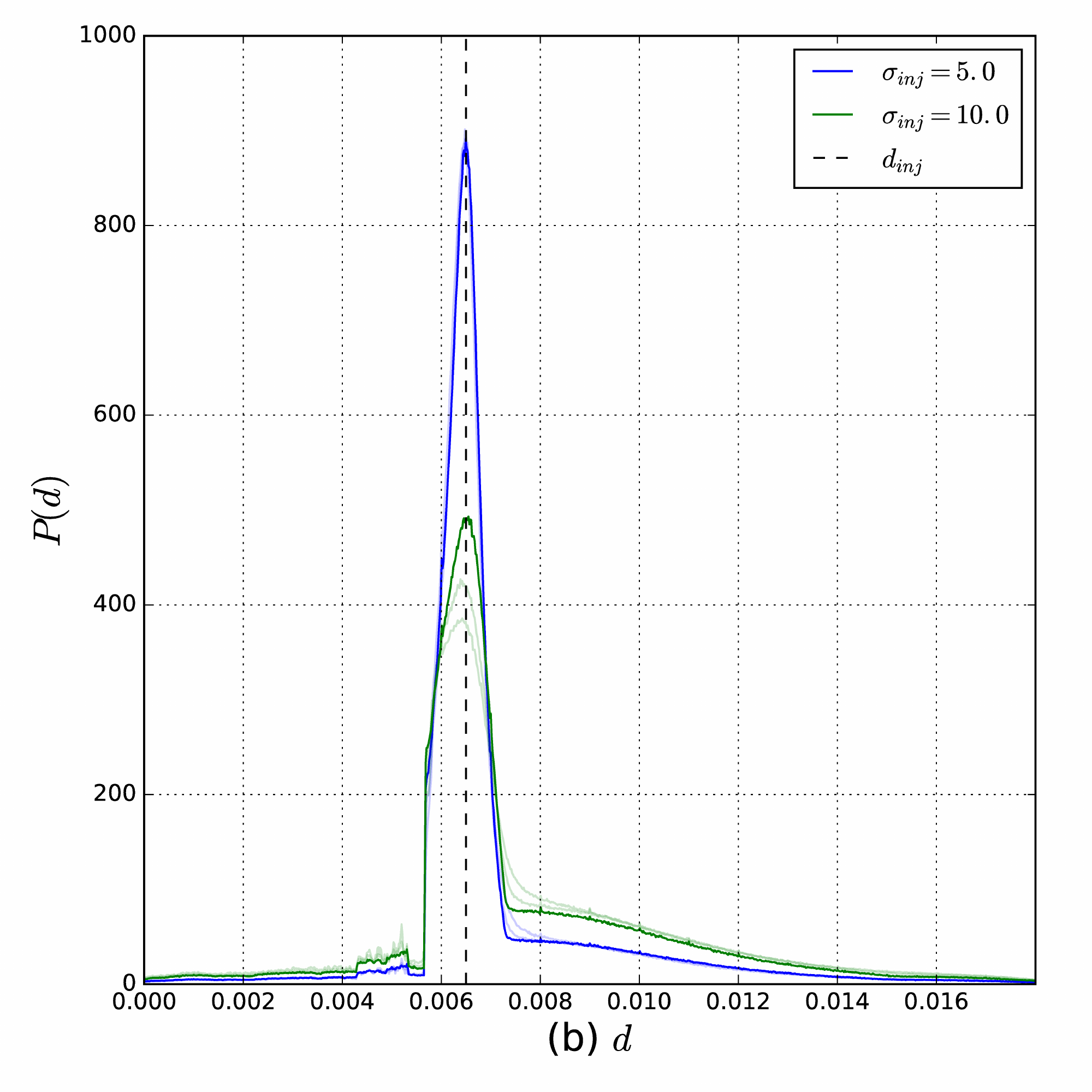}
\includegraphics[width=0.32\linewidth]{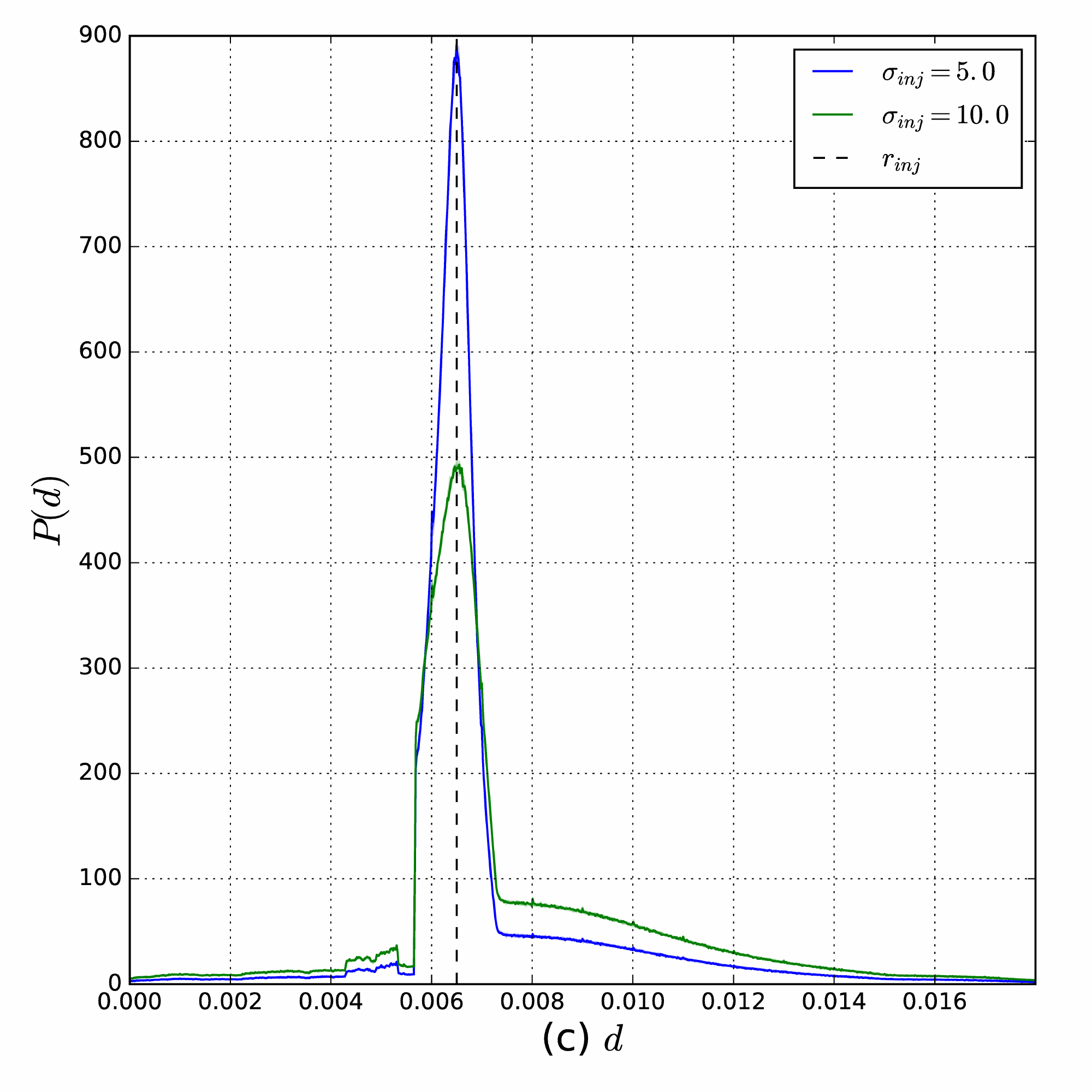}
\includegraphics[width=0.32\linewidth]{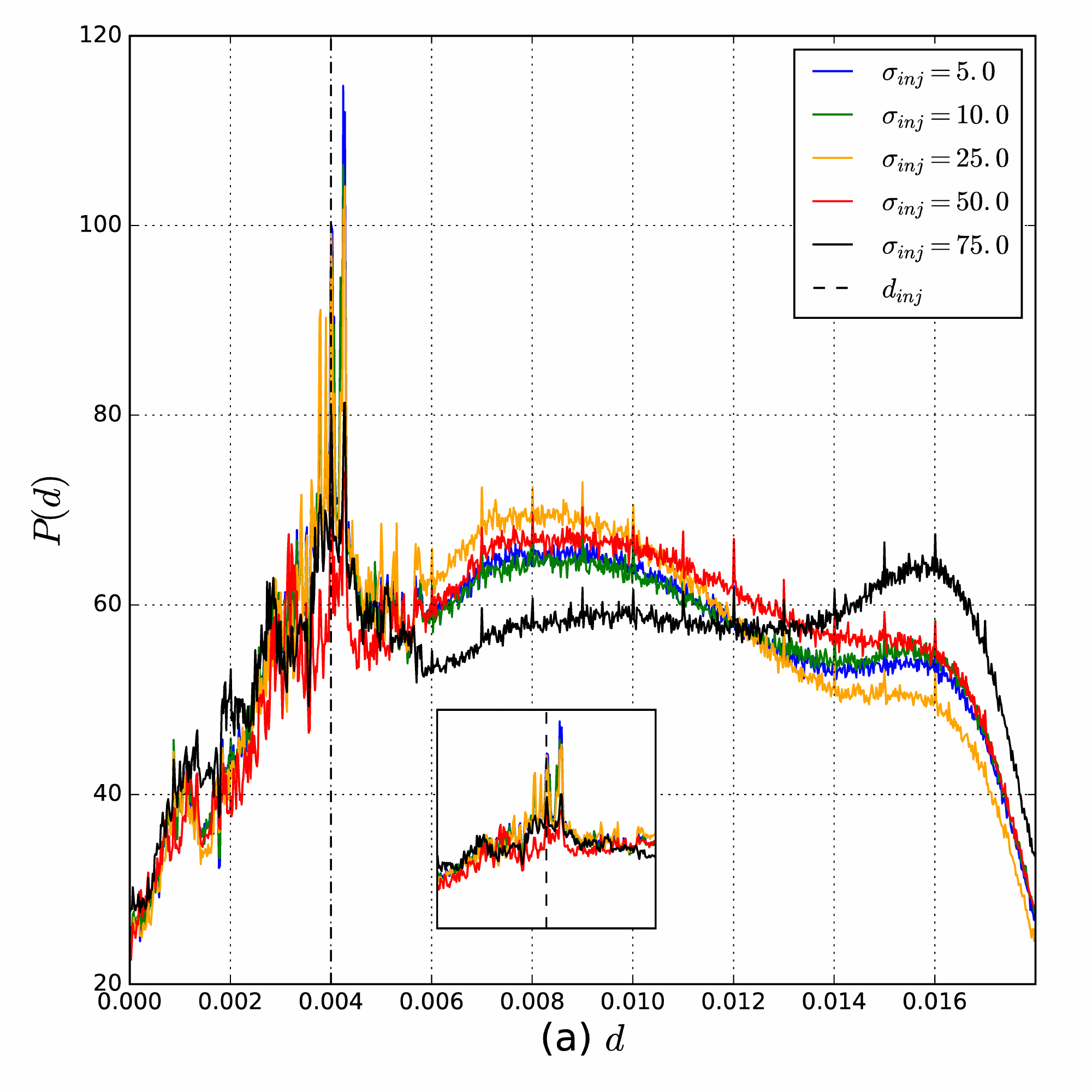}
\includegraphics[width=0.32\linewidth]{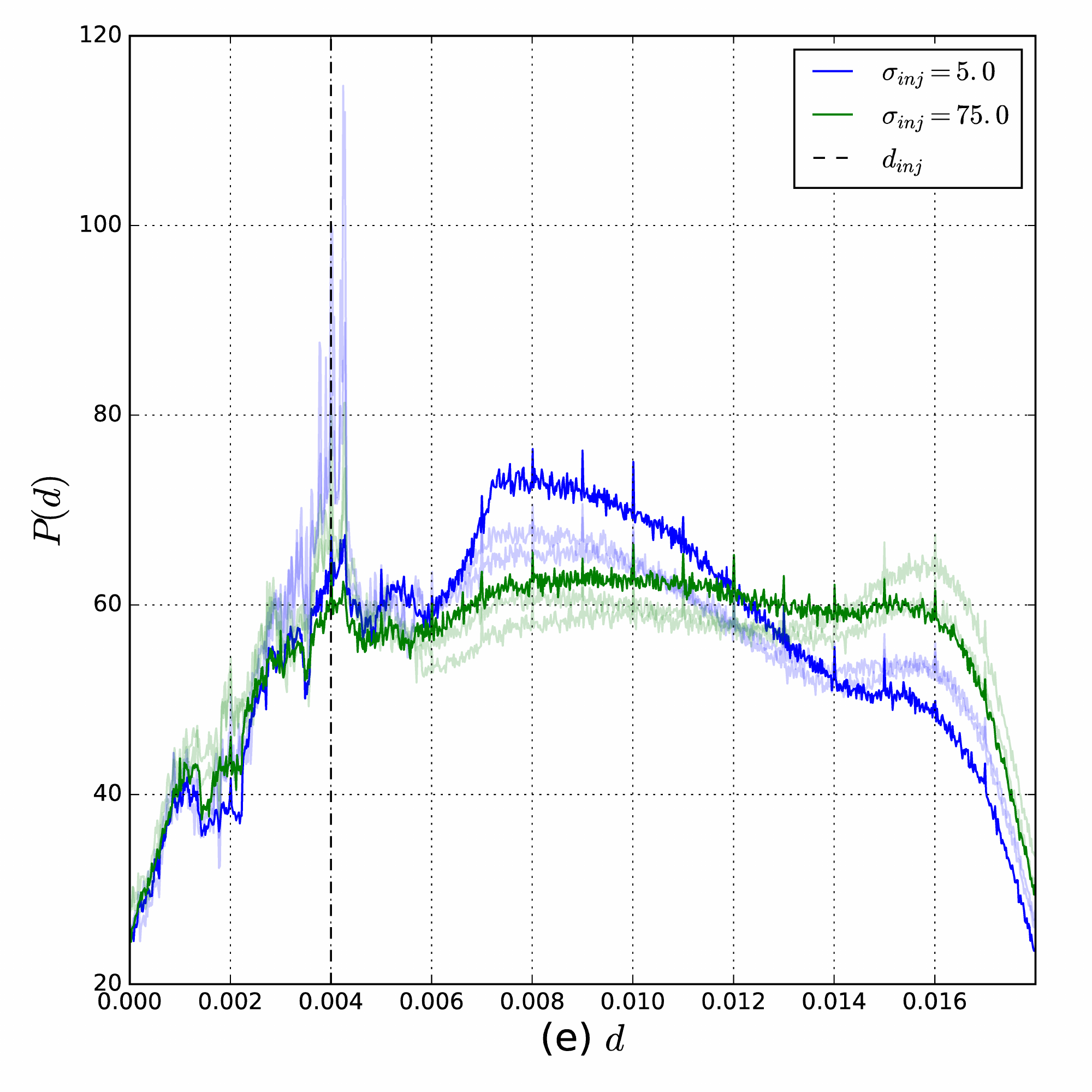}
\includegraphics[width=0.32\linewidth]{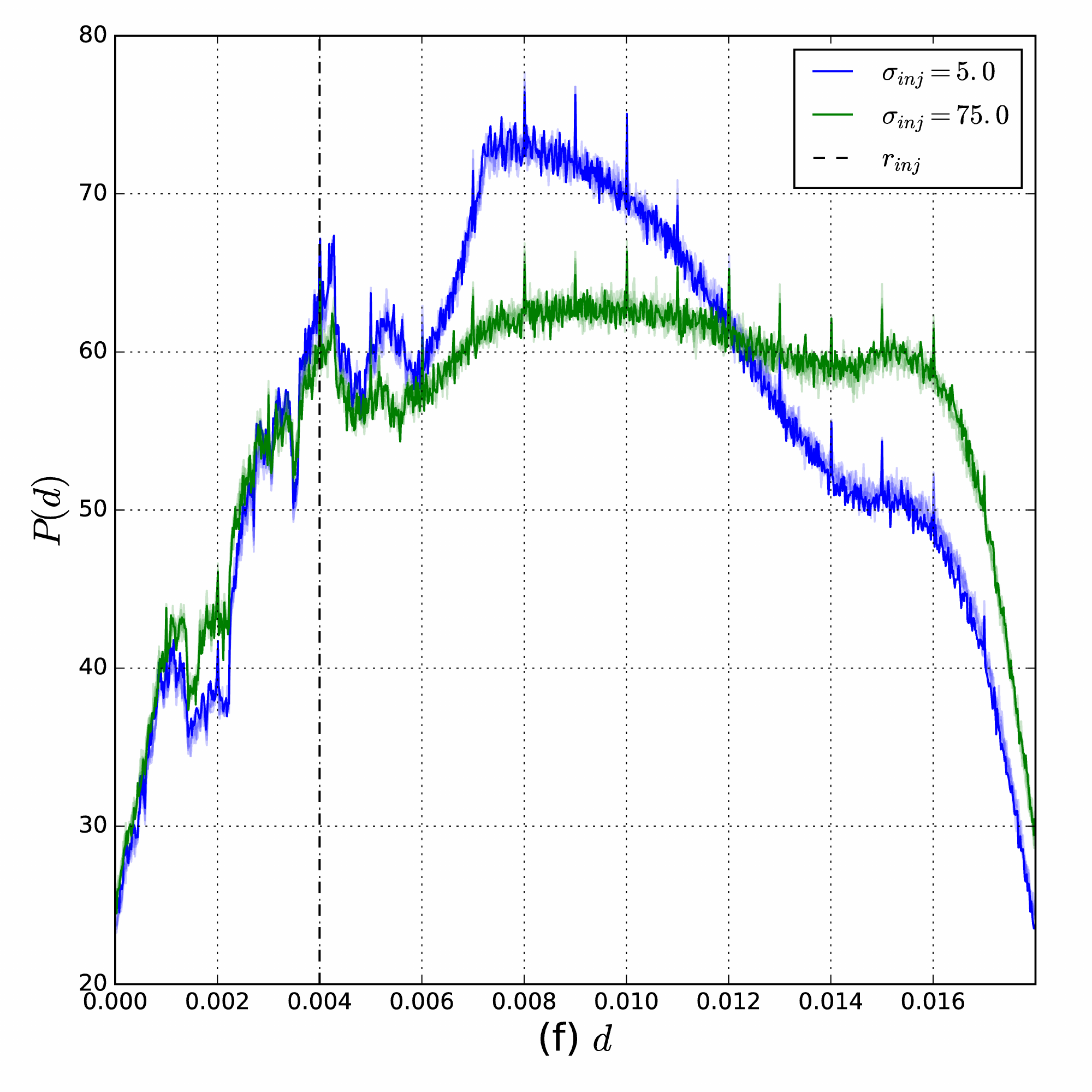}
\caption{\label{fig:droso_param_dependence}\emph{Top Row}: Posterior probability distributions for periodic data, $p(d | \mathcal{Y}_p)$. \emph{Bottom Row}: Posterior probability distributions for chaotic data, $p(d | \mathcal{Y}_c)$. \emph{Left Panel}: Posterior probability distributions for different levels of injected noise, $\sigma_{inj}$ = 5 (blue), 10 (green), 25 (orange), 50 (red), 75 (blue). \emph{Middle Panel}: Posterior probability distributions for different lengths of data sets, $t=\{50,100,500\}$ for a given injected noise $\sigma_{inj}$ = 5 (blue) and 10 (green) (for periodic data (b)), and $\sigma_{inj}$ = 5 (blue) and 75 (green) (for chaotic data (e)). The $t=500$ is solid, whereas the other curves are translucent. \emph{Right Panel}: Posterior probability distributions for different levels of recovered noise, $t=\{5, 10, 25, 50, 75\}$ for a given injected noise $\sigma_{inj}$ = 5 (blue) and 10 (green) (for periodic data (c)), and $\sigma_{inj}$ = 5 (blue) and 75 (green) (for chaotic data (f)). The $t=500$ is solid, whereas the other curves are translucent. The inset in (d) shows a close up around the injected value of $0.004$ for chaotic data, giving rise to a secondary mode.}
\end{figure*}

\begin{figure}[t!]
\centering
\includegraphics[width=\linewidth]{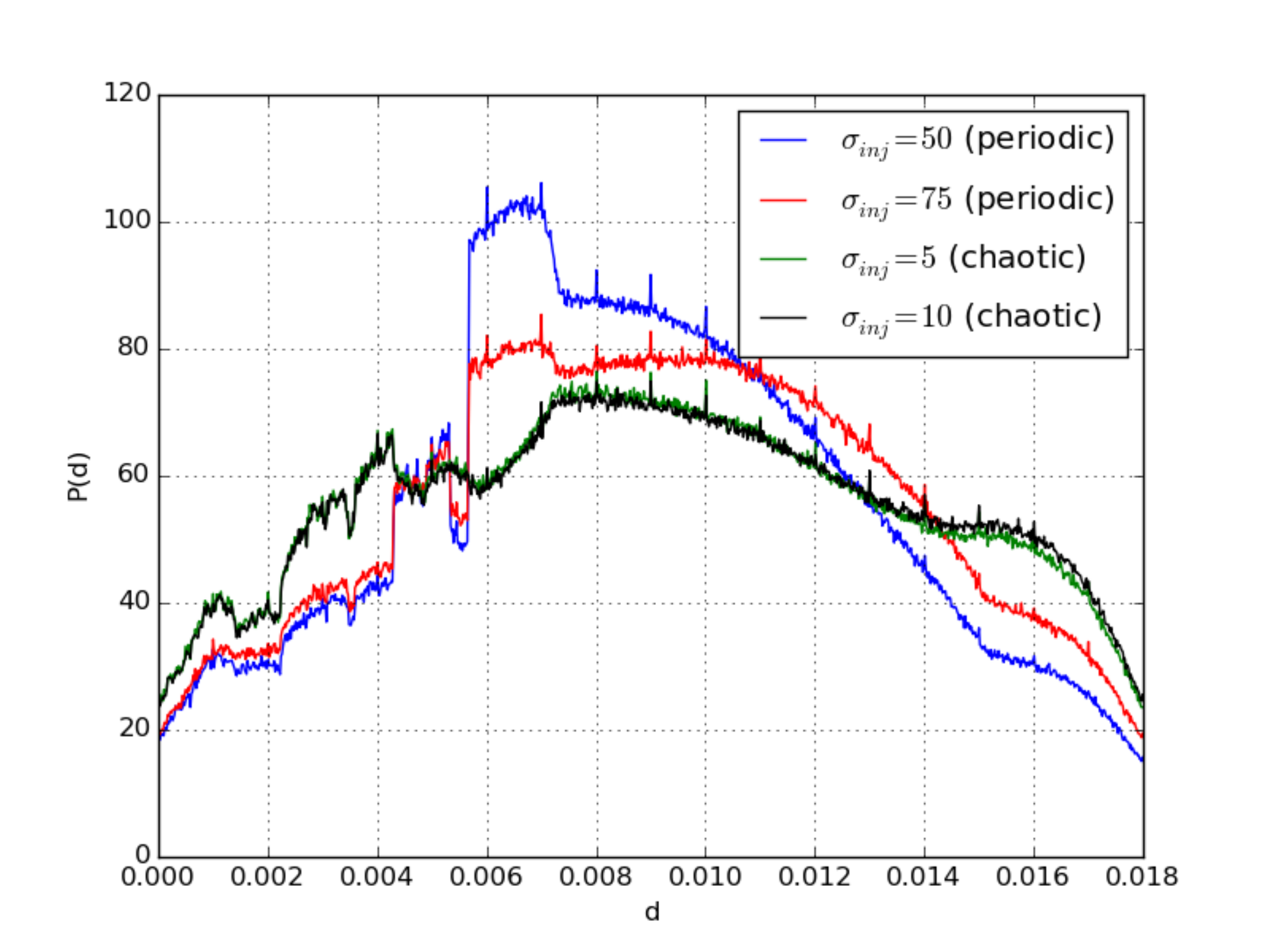}
\caption{\label{fig:droso_pe_vs_ch}Posterior probability distributions $p(d | \mathcal{Y}$ for periodic and chaotic data.}
\end{figure}

The Metropolis Hastings algorithm shows the same robustness in recovering the injected value of $d$, as in the case of the logistic map (Fig.~\ref{fig:robustness_droso}). Also, the posterior probability distribution of $d$, $p(d |\mathcal{Y},I)$ depends on nature of data (periodic or chaotic), standard deviation of injected noise $\sigma_{inj}$, length of the data set $d$ , and the standard deviation of the recovered noise $\sigma_r$ (Fig.~\ref{fig:droso_param_dependence}), in a manner similar to the general logistic map model. Another result qualitatively similar to the logistic map case is shown in Fig.~\ref{fig:droso_pe_vs_ch}, where as $\sigma _{inj}$ is increased, parameter estimation on periodic data becomes indistinguishable from parameter estimation on chaotic data.  

The difference between the precision of the parameter estimation, as assessed by the standard deviation of the posterior, can be most easily seen from the bottom panel of figure~\ref{fig:bd_d} which shows the so-called ``violin plot'' \cite{violin}. In particular, the x-axis shows the injected value and the y-axis shows the mean (in red dot) of the posterior, which is shows as the shaded region. We can clearly see that for parameter values with periodic dynamics, the posterior is peaked near the injected value while in the chaotic region, it is quite broad and even multi-modal. (The mean deviates from the injected value for the values near the boundary $d=0.018$ is that the prior is uniform only up to $d=0.018$, thus leading to ``aliasing'' effect.)

In particular, we would like to stress a few general conclusions. The posterior distribution in the case of chaotic data is multimodal, and neither the mean nor the mode give a good estimate of the parameter. Further, the short time series with $t=50$ points shows a prominent mode near the injected value which is much less clear in the case of the long time series with $t=500$ points, indicating that short time series is better at recovering the parameter value in case of chaotic data. This is qualitatively consistent with the remarks made in \cite{Hung-MIT-report} relating parameter estimation and shadowing - shadowing short trajectories is easier than long ones in chaotic systems. In either case, a qualitative interpretation of the posterior (for example, choice of mode rather than mean in case of short trajectory) will help choose a narrow prior for subsequent parameter estimation studies that progressively may help us identify the true injected parameter. Use of a long time series all at once is sub-optimal and more precise measurements (decreased $\sigma_{inj}$) has little effect, in the case of chaotic time series. For periodic time series, the longer and more precise measurements help identify the parameter with more precision and accuracy \cite{accuprec}.

%============================
% FIG 8: Robustness of MH algorithm
%============================
%-----------------------------

%============================
% FIG 9: droso parameter dependence
%============================
%-----------------------------

%============================
% FIG 10: periodic vs chaotic posteriors
%============================
%-----------------------------

\subsection{Estimation of two parameters}

\begin{figure*}[t!]
\centering
\includegraphics[width=0.45\textwidth]{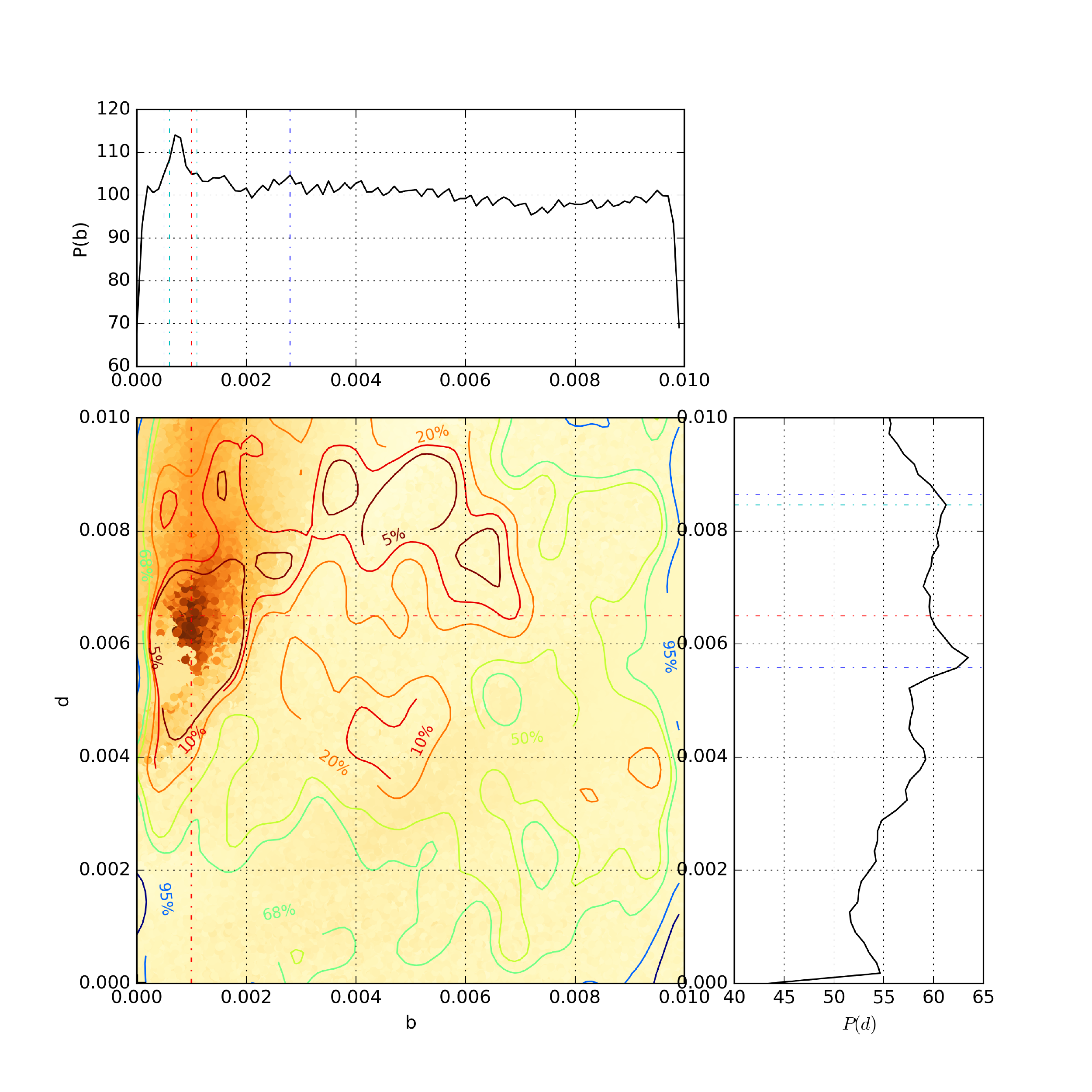}
\includegraphics[width=0.45\textwidth]{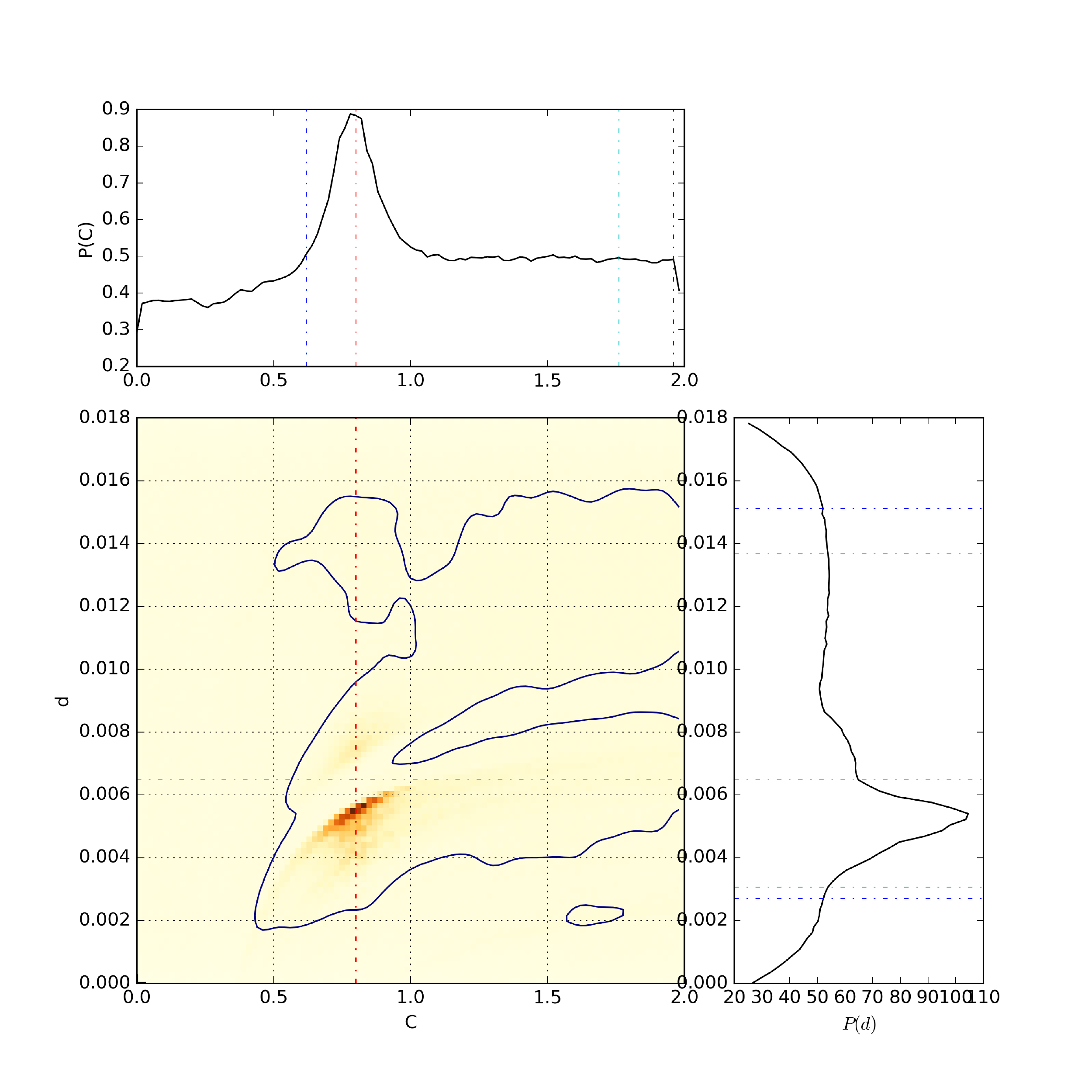}
\caption{\label{fig:nonlin-corr} Posterior probability distributions $p(b, d | \mathcal{Y})$ (left panel) and $p(C, d | \mathcal{Y})$ (right panel) for periodic data. The color bar shows the negative of the log posterior, and the red dashed lines indicate the injection values for the parameters.}
\end{figure*}

One of the main motivations for the use of Bayesian probabilistic methods of parameter estimation is to study the nonlinear inter-dependencies between different parameters. We illustrate this by trying to estimate two parameters instead of one using the Metropolis-Hastings algorithm for sampling the two parameter posterior distribution. We choose the parameter pair $\theta = (b, d)$, while the remaining parameters $\{C,c,f_{max},A,a\}$ are fixed at $\{$0.8, 0.002, 80.429, exp(0.592), 0.003$\}$. We perform the parameter estimation on the periodic data set described in Fig.~\ref{fig:data_droso}(a), with a length of 100 generations, and as earlier, a uniform prior. The noise injected into the data set is chosen to be low ($\sigma_{inj} = 5$) so as to give us the best possible chance of a good recovery. Fig.~\ref{fig:nonlin-corr} shows the 2-dimensinal posterior distribution of $(b, d)$. For the low noise periodic data case, the injection values can be recovered at even a $2\%$ confidence interval. The right panel of Fig.~\ref{fig:nonlin-corr} shows the results for the case of estimation of another set of parameters $\theta = (C,d)$ while keeping remaining parameters fixed. From these figures we see that there is significant nonlinear inter-dependence between the various parameters. Thus even though the multi-parameter estimation is significantly more imprecise, the information we gain about these inter-dependence will be lost if we only perform a single parameter estimation.

%============================
% FIG 11: 2D PE
%============================
%-----------------------------

\section{Conclusion and discussions} \label{sec:discuss}

In this paper, we investigate parameter estimation for chaotic dynamical systems using Bayesian framework. In particular, we study the qualitative properties of the posterior for the parameters to be estimated conditioned on observations of the system for two population dynamical models of increasing biological and dynamic complexity. Our main interest is to understand how well the parameter estimation works when there is little or incomplete prior information about the parameters (as is often the case in many ecological modeling scenarios). This is quantified by specifying an uninformative uniform prior in the Bayesian setup. We particularly focus on understanding the effect of the length of observational data, the actual (``injected'') noise in the data, and the representation of this noise in the posterior (``recovery'' noise).

A major aspect of our study is the careful distinction between (i) the actual observational errors that may be present in the data,
and (ii) the representation of these errors in the Bayesian parameter estimation. We show that the estimation is in fact quite sensitive to
the former (i.e. the observational uncertainties) but not much to the latter (i.e. the representation of these uncertainties). More specifically, in the case when the system exhibits periodic behavior, a large noise in the experimental data leads to very imprecise estimation of parameters even when the number of observations is large while a small noise leads to increasingly more accurate estimates with increasingly larger number of observations. In the case when the system exhibits chaotic behavior, the estimation remains imprecise, largely independent of the noise in the data or the number of observations, partly because of the phenomena discussed in detail in \cite{Hung-MIT-report} which relates it to shadowing properties of chaotic maps and also partly because of the statistical similarities between chaotic time series and noise. The above general remarks are mostly independent of the representation of the noise used
in the parameter estimation procedure.

The main conclusions can be summarized as follows.

\begin{enumerate}
\item The Metropolis-Hastings algorithm seems to be extremely robust in a one-parameter estimation problem, with the injection value of the parameter being almost always within a 68$\%$ confidence level of the posterior probability distribution.
\item The problem of parameter estimation is significantly more difficult for chaotic data, than periodic data. Even long and accurate time series of measurements do not lead to a accurate estimation of parameters in case the underlying dynamics is chaotic. Thus, prior knowledge about the system, quantified in terms of a narrow prior distribution, will be necessary for accurate and/or precise parameter estimation.
\item In contrast to the above, when the underlying dynamics is periodic, the parameter estimates become more precise and accurate as one increases the precision of the observations (i.e. decreasing the observational noise) and is only mildly affected by the length of observed trajectory.
\item The intrinsic noise in the experimental data seems to have the largest effect in the parameter estimation, while the effects of the data length and recovered noise is negligible. In the analysis performed above, it might seem that for the same level of experimental noise, the parameter estimation is better for shorter data sets, in case the underlying dynamics is chaotic. However more runs on simulated and real data needs to be performed before any firm conclusions can be drawn. The relation of this phenomena to ideas about shadowing \cite{Hung-MIT-report} will be investigated in future. Indeed, if the above conclusion is tenable, then it will indicate that for chaotic systems, precise measurements of even a short trajectory are more important than imprecise measurements of long trajectories.
\end{enumerate}
Our results are shown for the one-parameter, one variable discrete logistic model and a more complex but realistic two variable multi-parameter discrete model describing insect population growth. All the features discussed above are important in the context of ecological and epidemiological time series data, which are used for the assessment and prediction of population health. It is difficult to collect large data-sets - both in laboratory and in field surveys; observations include variations arising from multiple extrinsic and intrinsic sources leading to noisy data sets; model parameters can be correlated, and different kinds of dynamics (stable, periodic and complex) are observed in growth of populations of different organisms.

There are several avenues for further theoretical investigations of the parameter estimation problem. It would be interesting to investigate the effects of injected noise and the length of data on estimating multiple parameters, in both the periodic and chaotic dynamics. Sequential algorithms (for example, approximate ones, such as the Kalman or particle filter \cite{annan2004efficient, kantas2015particle}) which would use the posterior conditioned on a short trajectory as a prior for estimating parameters based on further data will certainly lead to improved estimation and such algorithms need to be investigated, in particular, for chaotic systems.

\section*{Data, code and materials}
Our paper has no data or associated material. All the codes use Python2.7, which is open source.

\section*{Competing interests}
We have no competing interests.

\section*{Authors' contributions}
Conceived and designed the simulations: AA, AG, SS. Performed mathematical analysis: AG, SB. Analyzed the simulations: AG, AA.  Wrote the paper: AG, AA, SS. All authors gave final approval for publication.

\section*{Acknowledgements}
SS acknowledges support from SERB as J C Bose Fellowship. Computations were performed at the ICTS cluster Mowgli.

\section*{Funding}
There is no funding to report for this submission.

\bibliography{drosorefs}

\begin{thebibliography}{10}

\bibitem{aguirre2009modeling}
Aguirre LA, Letellier C.
\newblock Modeling nonlinear dynamics and chaos: a review.
\newblock Mathematical Problems in Engineering. 2009;2009.

\bibitem{hong2008model}
Hong X, Mitchell RJ, Chen S, Harris CJ, Li K, Irwin GW.
\newblock Model selection approaches for non-linear system identification: a
  review.
\newblock International journal of systems science. 2008;39(10):925--946.

\bibitem{van2013detection}
Van~Trees HL, Bell KL.
\newblock Detection estimation and modulation theory, pt. I.
\newblock Wiley; 2013.

\bibitem{schafer2015framework}
Schafer CM.
\newblock A framework for statistical inference in astrophysics.
\newblock Annual Review of Statistics and Its Application. 2015;2:141--162.

\bibitem{abbott2016properties}
Abbott B, Abbott R, Abbott T, Abernathy M, Acernese F, Ackley K, et~al.
\newblock Properties of the binary black hole merger GW150914.
\newblock Physical review letters. 2016;116(24):241102.

\bibitem{ghosh2016testing}
Ghosh A, Ghosh A, Johnson-McDaniel NK, Mishra CK, Ajith P, Del~Pozzo W, et~al.
\newblock Testing general relativity using golden black-hole binaries.
\newblock Physical Review D. 2016;94(2):021101.

\bibitem{RobertC99}
Robert CP, Casella G.
\newblock {M}onte {C}arlo statistical methods.
\newblock New York: Springer; 1999.

\bibitem{JASA93_443p1032}
Liu JS, Chen R.
\newblock Sequential Monte Carlo Methods for Dynamic Systems.
\newblock Journal of the American Statistical Association. 1998;93:1032--1044.

\bibitem{ApteH07}
Apte A, Hairer M, Stuart AM, Voss J.
\newblock Sampling the posterior: An approach to non-Gaussian data
  assimilation.
\newblock Physica D. 2007;230:50--64.

\bibitem{ApteJ08}
Apte A, Jones CKRT, Stuart AM, Voss J.
\newblock Data assimilation: Mathematical and statistical perspectives.
\newblock Int~J~Numerical Methods in Fluids. 2008;56:1033--1046.

\bibitem{kantas2015particle}
Kantas N, Doucet A, Singh SS, Maciejowski J, Chopin N, et~al.
\newblock On particle methods for parameter estimation in state-space models.
\newblock Statistical science. 2015;30(3):328--351.

\bibitem{glass1988clocks}
Glass L, Mackey MC.
\newblock From clocks to chaos: the rhythms of life.
\newblock Princeton University Press; 1988.

\bibitem{suguna1999minimal}
Suguna C, Chowdhury KK, Sinha S.
\newblock Minimal model for complex dynamics in cellular processes.
\newblock Physical Review E. 1999;60(5):5943.

\bibitem{singh2004role}
Singh BK, Rao JS, Ramaswamy R, Sinha S.
\newblock The role of heterogeneity on the spatiotemporal dynamics of
  host--parasite metapopulation.
\newblock Ecological modelling. 2004;180(2):435--443.

\bibitem{coelho2011bayesian}
Coelho FC, Code{\c{c}}o CT, Gomes MGM.
\newblock A bayesian framework for parameter estimation in dynamical models.
\newblock PloS one. 2011;6(5):e19616.

\bibitem{alkema2008bayesian}
Alkema L, Raftery AE, Brown T.
\newblock Bayesian melding for estimating uncertainty in national HIV
  prevalence estimates.
\newblock Sexually transmitted infections. 2008;84(Suppl 1):i11--i16.

\bibitem{bettencourt2008real}
Bettencourt LM, Ribeiro RM.
\newblock Real time bayesian estimation of the epidemic potential of emerging
  infectious diseases.
\newblock PLoS One. 2008;3(5):e2185.

\bibitem{calderhead2009accelerating}
Calderhead B, Girolami M, Lawrence ND.
\newblock Accelerating Bayesian inference over nonlinear differential equations
  with Gaussian processes.
\newblock In: Advances in neural information processing systems; 2009. p.
  217--224.

\bibitem{vyshemirsky2008biobayes}
Vyshemirsky V, Girolami M.
\newblock BioBayes: a software package for Bayesian inference in systems
  biology.
\newblock Bioinformatics. 2008;24(17):1933--1934.

\bibitem{golightly2008bayesian}
Golightly A, Wilkinson DJ.
\newblock Bayesian inference for nonlinear multivariate diffusion models
  observed with error.
\newblock Computational Statistics \& Data Analysis. 2008;52(3):1674--1693.

\bibitem{gao2012bayesian}
Gao M, Chang X, Wang X.
\newblock Bayesian parameter estimation in dynamic population model via
  particle Markov chain Monte Carlo.
\newblock Computational Ecology and Software. 2012;2(4):181.

\bibitem{de2002fitting}
de~Valpine P, Hastings A.
\newblock Fitting population models incorporating process noise and observation
  error.
\newblock Ecological Monographs. 2002;72(1):57--76.

\bibitem{rasmussen2011inference}
Rasmussen DA, Ratmann O, Koelle K.
\newblock Inference for nonlinear epidemiological models using genealogies and
  time series.
\newblock PLoS Comput Biol. 2011;7(8):e1002136.

\bibitem{harmon1997markov}
Harmon R, Challenor P.
\newblock A Markov chain Monte Carlo method for estimation and assimilation
  into models.
\newblock Ecological modelling. 1997;101(1):41--59.

\bibitem{calder2003incorporating}
Calder C, Lavine M, M{\"u}ller P, Clark JS.
\newblock Incorporating multiple sources of stochasticity into dynamic
  population models.
\newblock Ecology. 2003;84(6):1395--1402.

\bibitem{andrieu2010particle}
Andrieu C, Doucet A, Holenstein R.
\newblock Particle markov chain monte carlo methods.
\newblock Journal of the Royal Statistical Society: Series B (Statistical
  Methodology). 2010;72(3):269--342.

\bibitem{Hung-MIT-report}
Hung ES.
\newblock Parameter Estimation in Chaotic Systems.
\newblock MIT Artificial Intelligence Laboratory; 1995. AITR-1541.

\bibitem{baker1996inverting}
Baker GL, Gollub JP, Blackburn JA.
\newblock Inverting chaos: Extracting system parameters from experimental data.
\newblock Chaos: An Interdisciplinary Journal of Nonlinear Science.
  1996;6(4):528--533.

\bibitem{kostelich1992problems}
Kostelich EJ.
\newblock Problems in estimating dynamics from data.
\newblock Physica D: Nonlinear Phenomena. 1992;58(1-4):138--152.

\bibitem{jaeger1996unbiased}
Jaeger L, Kantz H.
\newblock Unbiased reconstruction of the dynamics underlying a noisy chaotic
  time series.
\newblock Chaos: An Interdisciplinary Journal of Nonlinear Science.
  1996;6(3):440--450.

\bibitem{voss2004nonlinear}
Voss HU, Timmer J, Kurths J.
\newblock Nonlinear dynamical system identification from uncertain and indirect
  measurements.
\newblock International Journal of Bifurcation and Chaos.
  2004;14(06):1905--1933.

\bibitem{abarbanel2009dynamical}
Abarbanel HD, Creveling DR, Farsian R, Kostuk M.
\newblock Dynamical state and parameter estimation.
\newblock SIAM Journal on Applied Dynamical Systems. 2009;8(4):1341--1381.

\bibitem{alonso2015parameter}
Alonso LM.
\newblock Parameter estimation, nonlinearity, and Occam's razor.
\newblock Chaos: An Interdisciplinary Journal of Nonlinear Science.
  2015;25(3):033104.

\bibitem{sorrentino2009using}
Sorrentino F, Ott E.
\newblock Using synchronization of chaos to identify the dynamics of unknown
  systems.
\newblock Chaos: An Interdisciplinary Journal of Nonlinear Science.
  2009;19(3):033108.

\bibitem{parlitz1996synchronization}
Parlitz U, Junge L, Kocarev L.
\newblock Synchronization-based parameter estimation from time series.
\newblock Physical Review E. 1996;54(6):6253.

\bibitem{parlitz1996estimating}
Parlitz U.
\newblock Estimating model parameters from time series by autosynchronization.
\newblock Physical Review Letters. 1996;76(8):1232.

\bibitem{maybhate1999use}
Maybhate A, Amritkar R.
\newblock Use of synchronization and adaptive control in parameter estimation
  from a time series.
\newblock Physical Review E. 1999;59(1):284.

\bibitem{maybhate2000dynamic}
Maybhate A, Amritkar R.
\newblock Dynamic algorithm for parameter estimation and its applications.
\newblock Physical Review E. 2000;61(6):6461.

\bibitem{amritkar2009estimating}
Amritkar R.
\newblock Estimating parameters of a nonlinear dynamical system.
\newblock Physical Review E. 2009;80(4):047202.

\bibitem{sugihara1990distinguishing}
Sugihara G, Grenfell B, May RM, Chesson P, Platt H, Williamson M.
\newblock Distinguishing error from chaos in ecological time series [and
  discussion].
\newblock Philosophical Transactions of the Royal Society of London B:
  Biological Sciences. 1990;330(1257):235--251.

\bibitem{may1976simple}
May RM, et~al.
\newblock Simple mathematical models with very complicated dynamics.
\newblock Nature. 1976;261(5560):459--467.

\bibitem{meyer2000bayesian}
Meyer R, Christensen N.
\newblock Bayesian reconstruction of chaotic dynamical systems.
\newblock Physical Review E. 2000;62(3):3535.

\bibitem{meyer2001fast}
Meyer R, Christensen N.
\newblock Fast Bayesian reconstruction of chaotic dynamical systems via
  extended Kalman filtering.
\newblock Physical Review E. 2001;65(1):016206.

\bibitem{mcsharry1999better}
McSharry PE, Smith LA.
\newblock Better nonlinear models from noisy data: Attractors with maximum
  likelihood.
\newblock Physical Review Letters. 1999;83(21):4285.

\bibitem{ghasemi2011bayesian}
Ghasemi O, Lindsey ML, Yang T, Nguyen N, Huang Y, Jin YF.
\newblock Bayesian parameter estimation for nonlinear modelling of biological
  pathways.
\newblock BMC systems biology. 2011;5(Suppl 3):S9.

\bibitem{liepe2014framework}
Liepe J, Kirk P, Filippi S, Toni T, Barnes CP, Stumpf MP.
\newblock A framework for parameter estimation and model selection from
  experimental data in systems biology using approximate Bayesian computation.
\newblock Nature protocols. 2014;9(2):439--456.

\bibitem{kruschke2014doing}
Kruschke J.
\newblock Doing Bayesian data analysis: A tutorial with R, JAGS, and Stan.
\newblock Academic Press; 2014.

\bibitem{robert2013monte}
Robert C, Casella G.
\newblock Monte Carlo statistical methods.
\newblock Springer Science \& Business Media; 2013.

\bibitem{reich2015probabilistic}
Reich S, Cotter C.
\newblock Probabilistic forecasting and Bayesian data assimilation.
\newblock Cambridge University Press; 2015.

\bibitem{KalnayBook}
Kalnay E.
\newblock Atmospheric modeling, data assimilation and predictability.
\newblock Cambridge University Press; 2003.

\bibitem{feigenbaum1978quantitative}
Feigenbaum MJ.
\newblock Quantitative universality for a class of nonlinear transformations.
\newblock Journal of statistical physics. 1978;19(1):25--52.

\bibitem{10.2307/2461536}
Timothy~Prout FM.
\newblock Competition Among Immatures Affects Their Adult Fertility: Population
  Dynamics.
\newblock The American Naturalist. 1985;126(4):521--558.
\newblock Available from: \url{http://www.jstor.org/stable/2461536}.

\bibitem{doi:10.1086/284890}
Mueller LD.
\newblock Density-Dependent Population Growth and Natural Selection in
  Food-Limited Environments: The Drosophila Model.
\newblock The American Naturalist. 1988;132(6):786--809.
\newblock Available from: \url{http://dx.doi.org/10.1086/284890}.

\bibitem{Mueller01061988}
Mueller LD.
\newblock Evolution of competitive ability in Drosophila by density-dependent
  natural selection.
\newblock Proceedings of the National Academy of Sciences.
  1988;85(12):4383--4386.
\newblock Available from:
  \url{http://www.pnas.org/content/85/12/4383.abstract}.

\bibitem{10.2307/4835}
Rodriguez DJ.
\newblock A Model of Population Dynamics for the Fruit Fly Drosophila
  melanogaster with Density Dependence in More than One Life Stage and Delayed
  Density Effects.
\newblock Journal of Animal Ecology. 1989;58(2):349--365.
\newblock Available from: \url{http://www.jstor.org/stable/4835}.

\bibitem{sinhajoshi}
Sinha S, Joshi A.
\newblock A life-stage based model of {D}rosophila melanogaster population
  growth in the laboratory;.
\newblock Under preparation.

\bibitem{violin}
Wikipedia. Violin plot --- Wikipedia{,} The Free Encyclopedia; 2016.
\newblock [Online; accessed 11-November-2016].
\newblock Available from:
  \url{https://en.wikipedia.org/w/index.php?title=Violin_plot&oldid=748930714}.

\bibitem{accuprec}
Wikipedia. Accuracy and precision --- Wikipedia{,} The Free Encyclopedia; 2016.
\newblock [Online; accessed 14-October-2016].
\newblock Available from:
  \url{https://en.wikipedia.org/w/index.php?title=Accuracy_and_precision&oldid=744242441}.

\bibitem{annan2004efficient}
Annan J, Hargreaves J.
\newblock Efficient parameter estimation for a highly chaotic system.
\newblock Tellus A. 2004;56(5):520--526.

\end{thebibliography}
\bibliographystyle{vancouver}

\end{document}